\documentclass[12pt]{article} 
\usepackage{graphicx,subfigure,latexsym}

\newcommand{\be}{\begin{equation}}
\newcommand{\ee}{\end{equation}}
\newcommand{\bear}{\begin{eqnarray}}
\newcommand{\eear}{\end{eqnarray}}
\newcommand{\ba}{\begin{array}}
\newcommand{\ea}{\end{array}}

 \newcommand{\CD}{{\cal D}}

\newcommand{\CR}{{\cal R}} \newcommand{\CW}{{\cal W}}

\begin{document}

\begin{titlepage}
\vfill
\begin{flushright}
{\normalsize arXiv:xxxx.xxxx[hep-th]}\\
\end{flushright}

\vfill
\begin{center}
{\Large\bf Anomalies and time reversal invariance }
\vskip0.3cm
{\Large\bf in relativistic hydrodynamics:}
\vskip0.3cm
{\Large\bf the second order and higher dimensional formulations}

\vskip 0.3in

Dmitri E. Kharzeev$^{1,2}$\footnote{e-mail:
{\tt dmitri.kharzeev@stonybrook.edu}}
 and 
Ho-Ung Yee$^{1}$\footnote{e-mail:
{\tt hyee@tonic.physics.sunysb.edu}}
\vskip 0.15in

 {\it $^{1}$Department of Physics and Astronomy, Stony Brook University,} \\
{\it Stony Brook, New York 11794-3800, USA }\\[0.3in]
{\it $^{2}$Department of Physics, Brookhaven National Laboratory,} \\
{\it Upton, New York 11973-5000, USA }\\[0.3in]

{\normalsize  May 30, 2011}

\end{center}

\vfill

\begin{abstract}
We present two new results on relativistic hydrodynamics with anomalies and external electromagnetic fields, ``Chiral MagnetoHydroDynamics" (CMHD). 
First, we study CMHD in four dimensions at second order in the derivative expansion  
assuming the conformal/Weyl invariance. We classify all possible independent conformal 
second order viscous corrections to the energy-momentum tensor and to the $U(1)$ current in the presence of external electric and/or magnetic fields, and identify eighteen terms that originate from the triangle anomaly.
We then propose and motivate the following guiding principle to constrain the CMHD: the anomaly--induced terms that are even under the time reversal invariance should not contribute to the local entropy production rate. This allows us to fix thirteen out of the eighteen transport coefficients that enter the second order formulation of CMHD.  
We also relate one of our second order transport coefficients to the chiral shear waves.
Our second subject is hydrodynamics with $(N+1)$-gon anomaly in an arbitrary $2N$ dimensions. 
The effects from the $(N+1)$-gon anomaly appear in hydrodynamics at $(N-1)$'th order in the derivative expansion,
and we identify precisely $N$ such corrections to the $U(1)$ current. The time reversal constraint 
is powerful enough to allow us to find the analytic expressions for all transport coefficients.
We confirm the validity of our results (and of the proposed guiding principle) by an explicit fluid/gravity computation within the AdS/CFT correspondence.

\end{abstract}

\vfill

\end{titlepage}
\setcounter{footnote}{0}

\baselineskip 18pt \pagebreak
\renewcommand{\thepage}{\arabic{page}}
\pagebreak

\section{Introduction and Summary}

 Quantum anomalies are among the most beautiful, subtle, and important phenomena in quantum field theory. Recently it has become clear that anomalies play a very important role also in the macroscopic dynamics of relativistic fluids. Much of this progress is motivated by the possibility to observe the anomalous "chiral magnetic" currents in non-Abelian quark-gluon plasma produced at Relativistic Heavy Ion Collider and the Large Hadron Collider. However, anomalous currents in hydrodynamics can be expected to have many other applications: for example, 
the time-reversal-invariant, non-dissipative currents in strongly correlated systems that we will discuss in this paper present clear interest for quantum computing. 

The Chiral Magnetic Effect (CME) is the anomaly--induced phenomenon of  electric charge separation along the axis of the applied magnetic field in the presence of fluctuating topological charge \cite{Kharzeev:2004ey,Kharzeev:2007tn,Kharzeev:2007jp,Fukushima:2008xe,Kharzeev:2009fn}. The CME in QCD coupled to electromagnetism assumes a chirality asymmetry
between left- and right-handed quarks, parameterized by a chiral 
chemical potential $\mu_A$.  Such an asymmetry can arise if there is
an asymmetry between the 
topology-changing transitions early in a heavy ion
collision.  
Closely related phenomena have been discussed in the physics of neutrinos  \cite{Vilenkin:1979ui}, 
primordial electroweak plasma \cite{Giovannini:1997gp} and quantum
wires \cite{acf}.  

While the original derivation used the weak coupling methods (see also Refs.\cite{Hou:2011ze,Landsteiner:2011cp} for recent discussions on this), the origin of the effect is essentially topological and so the CME is not renormalized even at strong coupling, as was shown by the holographic methods  \cite{Yee:2009vw,Rebhan:2009vc,Rubakov:2010qi,Gynther:2010ed,Gorsky:2010xu,Brits:2010pw,Kalaydzhyan:2011vx}. The evidence for the CME has been found in lattice QCD coupled to electromagnetism, both within the quenched approximation \cite{Buividovich:2009wi,Buividovich:2009zzb,Buividovich:2010tn} and with light domain wall fermions \cite{Abramczyk:2009gb}. Unlike the baryon chemical potential, the chiral chemical potential $\mu_A$ does not present a ``sign problem"  which opens a possibility for lattice computations at finite $\mu_A$ \cite{Fukushima:2008xe}. A direct lattice study of the chiral magnetic current as a function of $\mu_A$ was performed very recently \cite{Yamamoto:2011gk}; it confirms the expected dependence of CME on the chiral chemical potential and the magnetic field. 

Recently, STAR \cite{:2009uh, :2009txa} and PHENIX \cite{phenix,Ajitanand:2010rc}
Collaborations at Relativistic Heavy Ion Collider at BNL reported
experimental observation of charge asymmetry fluctuations.  While the
interpretation of the observed effect is still under intense
discussion, the fluctuations in charge asymmetry have been predicted \cite{Kharzeev:2004ey}
to occur in heavy ion collisions due to the CME. Very recently, STAR reported \cite{STAR-QM} the expected 
\cite{Kharzeev:2007jp,Fukushima:2008xe} disappearance of the effect at low collision energies where the energy density of created matter is smaller and likely below the critical one needed for the restoration of chiral symmetry. The ALICE Collaboration at the Large Hadron Collider at CERN has just reported \cite{ALICE-QM} the observation of charge asymmetry fluctuation signaling the persistence of the effect at very high energy densities.
Additional future tests of CME include the positive correlation between the electric and baryon charge asymmetries \cite{Kharzeev:2010gr}. See also Ref.\cite{KerenZur:2010zw}. 

Since in the strong coupling regime the plasma represents a fluid (for reviews, see \cite{Son:2007vk,Schafer:2009dj}), it is of great interest to study the effects 
of anomalies in relativistic hydrodynamics. 
A purely hydrodynamical derivation of the anomaly effects at first order in derivative expansion was given by Son and Surowka \cite{Son:2009tf}, motivated by earlier results in AdS/CFT correspondence  \cite{Erdmenger:2008rm,Banerjee:2008th,Torabian:2009qk} which found, among others, chiral vortical effect.
It has been generalized to anomalous superfluids \cite{Lublinsky:2009wr,Lin:2011mr,Bhattacharya:2011tr} and non-abelian symmetry \cite{Eling:2010hu,Neiman:2010zi}. It has been found that the CME current persists in hydrodynamics 
\cite{Sadofyev:2010pr} 
and is transferred by the sound-like gapless excitation -- ``the chiral magnetic wave" \cite{Kharzeev:2010gd,Burnier:2011bf}, see also \cite{Newman:2005hd} for an earlier study of collective excitations in anomalous hydrodynamics. 
Another line of development has been along the viewpoint of the effective field theory \cite{Sadofyev:2010is}.

The idea of Son and Surowka \cite{Son:2009tf} was to consider the local entropy production rate $\partial_\mu s^\mu$
and to impose on it the positivity constraint following from the second law of thermodynamics. The contributions from the anomaly to the entropy production 
were shown to be locally unbounded in either sign so that unless their coefficients identically vanished, they
could potentially violate the second law of thermodynamics. These arguments lead to a set of algebro-differential equations for the transport coefficients related to the anomaly; in many cases they can be solved.
 
The present work continues the investigation of anomalies in relativistic hydrodynamics. 
We also consider the entropy production as an important constraint. However, we propose a different 
guiding principle: instead of requiring the positivity of the total entropy production rate, we argue that the anomaly--induced viscous corrections should not contribute to the
entropy production at all, due to the time--reversal invariance of the anomalous transport coefficients. The time-reversal ${\cal T}$ invariance provides a unique criterion that can be used to establish the nature of currents. For example, the ``usual" electric conductivity $\sigma$ is ${\cal T}$-odd, as can be easily inferred from the Ohm's law $J^i = \sigma E^i$: the electric field is ${\cal T}$-even, whereas the electric current  $J^i$ is ${\cal T}$-odd. On the other hand,  the (anomalous) quantum Hall conductance  is a ${\cal T}$-even quantity, as it is associated with a ${\cal T}$-odd magnetic field. The physical meaning of ${\cal T}$ invariance of transport coefficients is quite simple: ${\cal T}$-odd conductivities describe dissipative currents, whereas ${\cal T}$-even conductivities describe {\it non-dissipative} currents. The anomaly-induced currents are protected by topology and are thus of non-disipative nature; as such, they do not contribute to the entropy production. We will discuss this in more detail in section \ref{time-reverse}; we will verify that our guiding principle leads to non-trivial relations among the transport coefficients that are obeyed by the available results of the explicit holographic computations in section \ref{hol_check}.

Let us add that enforcing a positivity constraint on the entropy production is not easy when one considers a second (or higher) order in the derivative expansion in hydrodynamics. Indeed, one first has to consider only a subspace of possible configurations with vanishing previous order contributions to
meaningfully discuss constraints at the second (or higher) order. This typically results in very few useful constraints on the second or higher order transport coefficients. However, the 
${\cal T}$--invariance guiding principle that we propose here provides a stronger constraint on the  anomaly--induced terms, and in some cases allows to evaluate them. 
Note that the first order hydrodynamics is known to have problems with causality and to be numerically unstable, so from a practical point of view the second order formulation of relativistic hydrodynamics with anomalies and external electric/magnetic fields is highly desirable; it is the topic of the present paper. 

The paper is organized as follows. 
In the first part (section \ref{part1}), we consider the anomaly--induced viscous
corrections at second order in the derivative expansion in four space-time dimensions.
As the number of possible independent terms increases drastically at the second and higher orders, we
assume the underlying conformal symmetry to constrain the problem.
We first classify all possible second order viscous corrections to the energy-momentum tensor and the $U(1)$ current, including also the non-anomalous terms, in the presence of external electric/magnetic fields. 
Let us mention that Refs.\cite{Erdmenger:2008rm,Banerjee:2008th} previously classified possible second order conformal viscous terms 
without including external electric/magnetic fields.
Then, by considering discrete symmetries of charge conjugation and parity, we select eighteen terms with transport coefficients that are necessarily linear in the anomaly coefficient. These transport coefficients are time-reversal ${\cal T}$ even, and
we demand that they do not contribute to the entropy production. This enables us to fix thirteen out of the eighteen anomaly--induced transport coefficients.  

Our results for the second order transport coefficients related to the triangle anomaly are new and can be expected to be universal.
Four transport coefficients of interest for us were previously computed in the holographic AdS/CFT approach using the fluid/gravity correspondence \cite{Erdmenger:2008rm,Banerjee:2008th}, which makes it possible to check and confirm some of our results. This test is quite non-trivial, and
we consider it as an important evidence for the validity of our proposed guiding principle and of the relations that we derive from it.
However we also identify many other second order transport coefficients coming from the anomaly that have not been computed in the AdS/CFT setup or in any other framework, and 
a more thorough fluid/gravity computations for them are certainly very desirable.

We also explain the phenomenon of the chiral shear wave, that is, a helicity--dependent shear mode dispersion relation due to triangle anomaly, in terms of one of the second order anomalous transport coefficients in the energy-momentum tensor. Chiral shear wave was first observed in Refs.\cite{Sahoo:2009yq,Matsuo:2009xn,Nakamura:2009tf} via linearized hydrodynamic
analysis in AdS/CFT correspondence. However its relation to viscous transport coefficients was unclear
and we close this remaining gap in the present paper.

In the second part of this work (section \ref{part2}), we consider the anomaly effects in  hydrodynamics in a higher $2N$ dimensional spacetime, where the theory has an underlying $(N+1)$-gon anomaly. 
By using the charge conjugation and parity symmetries, we show that the effects of the anomaly first appear
at $(N-1)$'th order in the derivative expansion in the $U(1)$/entropy currents, and we identify precisely $N$ such terms. Although these terms are of a very high order in derivatives, the time reversal invariance still dictates that they should not contribute to the entropy production, and this principle provides us
a sufficient set of constraints to determine these terms completely and derive for them analytic expressions.
We then confirm our results via the fluid/gravity correspondence in an AdS/CFT setup, corroborating our guiding principle.

\section{Second order relativistic conformal hydrodynamics \\ with triangle anomaly}\label{part1}

\subsection{A primer on the conformal/Weyl covariant formalism}

In this subsection we review the formalism of conformal hydrodynamics basing mainly on Refs.\cite{Baier:2007ix,Loganayagam:2008is}; this will also allow us to introduce the notation. 
Conformal hydrodynamics by definition is covariant under Weyl transformations,
\bear
g_{\mu\nu}(x)\to e^{-2 \phi(x) }g_{\mu\nu}(x)\quad,
\eear
where $\phi(x)$ is an arbitrary scalar function on the spacetime. In our convention,
a Weyl covariant tensor with Weyl weight $w$ transforms as
\bear
T^{\mu\nu\cdots}_{\alpha\beta\cdots}(x) \to e^{w \phi(x) }T^{\mu\nu\cdots}_{\alpha\beta\cdots}(x) \quad,
\eear
so the metric tensor has $w=-2$. Note that the upper indexed inverse metric $g^{\mu\nu}$ has $w=+2$ instead.
In short, the constraints from conformal symmetry on hydrodynamics are simple: {\it i)} the energy-momentum tensor
and the local symmetry currents that constitute the basic elements of hydrodynamics should be Weyl covariant; {\it ii)} the energy-momentum tensor should be traceless up to a local Weyl anomaly (which in most cases is a higher order effect 
in the derivative expansion scheme). 
One can easily derive Ward-type identities for the assumed Weyl invariance, for example as in Ref.\cite{Baier:2007ix}, and obtain the transformation properties of the energy-momentum tensor and the currents. The energy-momentum tensor $T^{\mu\nu}$ which is
traceless $T^\mu_{\,\,\mu}=0$ has a Weyl weight $w=+6$ and a current $j^\mu$ has $w=+4$.
In hydrodynamics, one is dealing with locally varying thermodynamic variables such as temperature $T$, pressure $p$, chemical
potentials $\mu$, local velocity $u^\mu$, etc, and writes the energy-momentum tensor and symmetry currents in terms of these variables -- the resulting expressions  
are the so-called constitutive relations. Imposing Weyl covariance on these expressions is a powerful constraint
that reduces much arbitrariness one might have without the conformal symmetry. In writing constitutive relations, one typically
invokes the derivative expansion scheme; the zeroth order expressions in the Landau frame are
\bear
T^{\mu\nu}&=&(\epsilon+p)u^\mu u^\nu +p g^{\mu\nu} = 4p u^\mu u^\nu +pg^{\mu\nu} +\cdots\quad,\\
j^\mu &=& n u^\mu +\cdots\quad,
\eear
where we have used the tracelessness condition $\epsilon=3p$ in the first line; $n$ is the charge density.
This gives us the Weyl weights of $p$ ($w=+4$), $n$ ($w=+3$), and $u^\mu$ ($w=+1$). The temperature $T$  naturally has $w=+1$ and
the Gibbs-Duhem relation $\epsilon+p=Ts+\mu n$ implies that the entropy density $s$ has $w=+3$ and the chemical potential $\mu$ has $w=+1$. The zeroth order expression for the entropy current $s^\mu=s u^\mu +\cdots $ implies that $s^\mu$ has $w=+4$.
\vskip0.3cm

A useful mnemonic used in Ref.\cite{Baier:2007ix} is that the Weyl weight of a Weyl covariant tensor is given by
\bear
w= [{\rm mass \,\,dimension}] + [\#\,\, {\rm of\,\,upper\,\,indices}]-[\# \,\,{\rm of\,\,lower\,\,indices}]\quad.
\eear
For example, an external gauge potential $A_\mu$ that couples to $j^\mu$ has $w=1-1=0$, which is also confirmed by
considering the Weyl invariance of the action 
\bear
S\sim \int d^4x \,\sqrt{-g} \,A_\mu j^\mu\quad.
\eear
Let us point out yet another useful nmemonic derived from holography. A tensor in the 4 dimensional conformal field theory
appears in the holographic 5 dimensions as a component of a 5 dimensional bulk field. The radial wavefunction  
of that particular component near the AdS boundary has a form $r^{-w}$ when the AdS metric is written as
${dr^2\over r^2}+r^2 (dx^\mu)^2$. One can identify $w$ in the exponent as the Weyl weight of the corresponding tensor.
For example, a bulk gauge field contains a current $j_\mu$ and an external gauge potential $A_\mu$ as
\bear
A^{\rm bulk}_\mu = {1\over r^2} j_\mu + A_\mu\quad;
\eear
this expression conforms to the fact that $j_\mu$ has $w=+2$ (recall that $j_\mu=g_{\mu\nu}j^\nu$) and $A_\mu$ has $w=0$. 
The 4 dimensional metric $g_{\mu\nu}$ is present in the bulk metric as
\bear
g^{\rm bulk}_{\mu\nu} = r^2 g_{\mu\nu}\quad,
\eear
which is also consistent with $w=-2$. This mnemonic is natural because the holographic versions of Weyl transformations
are in fact the coordinate reparameterizations $r\to e^{-\phi}r$.

 \vskip0.3cm

For higher derivative viscous terms, the construction of  Weyl covariant expressions out of derivatives becomes non-trivial,
simply because the ``ordinary" covariant derivatives $\nabla_\mu$ are not Weyl covariant. 
The Weyl covariant formalism from Ref.\cite{Loganayagam:2008is} that we will use is based on the idea of introducing Weyl covariant derivatives
$\CD_\mu$ which contain suitable Weyl connections to make them covariant under the Weyl transformations. 
This is in close analogy with the usual electromagnetic/gravitational covariant derivatives.
In the case of electromagnetism, the connection introduced is a new dynamical degree of freedom one has to add to the
theory, but in our case of hydrodynamics we do not want to introduce new degrees of freedom. Therefore, one needs to construct
a Weyl connection that transforms properly out of the already existing hydrodynamic variables; in our case it is the velocity field $u^\mu$.
The situation is quite similar to the metric Christoffel connection that is constructed out of the metric itself.
We refer the reader for details to Ref.\cite{Loganayagam:2008is}, and will outline only the main features of this approach. 
\vskip0.3cm
The Weyl covariant derivatives $\CD_\mu$ share almost all common properties with the covariant derivatives, including the chain rule. The $\CD_\mu T^{\nu\cdots}_{\alpha\cdots}$ has the same Weyl weight $w$ of the original $T^{\nu\cdots}_{\alpha\cdots}$ so that $\CD_\mu$ has $w=0$. Note that $\CD^\mu$ has $w=+2$. 
The important properties of $\CD_\mu$ for the purpose of hydrodynamics are 
\bear
u^\mu \CD_\mu u^\nu =0\quad,\quad \CD_\mu u^\mu =0\quad,\quad \CD_\mu g_{\nu\alpha}=0\quad,\quad \CD_\mu \epsilon^{\nu\alpha\beta\gamma}=0\quad,\label{prop}
\eear
where $\epsilon^{\nu\alpha\beta\gamma}$ is the totally anti-symmetric covariant tensor which is easily found to have $w=+4$.
To illustrate how $\CD_\mu$ acts, $\CD_\mu$ acting on a scalar $f$ of Weyl weight $w$ is
\bear
\CD_\mu f = \nabla_\mu f +w \CW_\mu\quad,
\eear
where $\CW_\mu$ is an analog of electromagnetic connection, but constructed out of $u^\mu$ as
\bear
\CW_\mu = u^\nu \nabla_\nu u_\mu -{(\nabla_\nu u^\nu)\over 3} u_\mu\quad,
\eear
and $w$ is acting as a charge in electromagnetic analogy. Note that $\CW_\mu$ is first order in the derivative. For tensors with indices, $\CD_\mu$ involves not only $w \CW_\mu$, but in addition more connections acting on tensor indices like metric connections. We will rarely need this detail, but
an important fact is that these connections are all linear in $\CW_\mu$ so that $\CD_\mu$ increases the number of derivatives by precisely one.
As expected, commutators of $\CD_\mu$ bring us several kinds of Weyl covariant curvature tensors.
The simplest one is
\bear
[\CD_\mu,\CD_\nu]f = w \CW_{\mu\nu}f\quad,\label{rel1}
\eear
where $\CW_{\mu\nu}=\partial_\mu \CW_\nu -\partial_\nu \CW_\mu$ is a new Weyl covariant tensor with weight $w=0$.  
Another example is
\bear
[\CD_\mu,\CD_\nu]V_\alpha = w \CW_{\mu\nu} V_\alpha - \CR_{\mu\nu\alpha}^{\quad\,\,\beta}V_\beta\quad,\label{rel2}
\eear
which includes a Weyl covariant cousin of the Riemann tensor; however its symmetry properties are slightly different \cite{Loganayagam:2008is}.

The main point is that the basic hydrodynamic equations are in fact Weyl covariant -- that is, one can show that
\bear
 \nabla_\mu T^{\mu\nu}=\CD_\mu T^{\mu\nu} \quad,\quad \nabla_\mu j^\mu = \CD_\mu j^\mu\quad,
\eear
precisely when $T^{\mu\nu}$ ($j^\mu$) has $w=+6$ ($w=+4$) and is traceless $T^\mu_{\,\,\mu}=0$. Therefore the Weyl covariant formalism can present a useful framework for the studies of conformal hydrodynamics.

\subsection{First order conformal hydrodynamics with anomaly}

We will begin by applying the Weyl formalism to the first order hydrodynamics with triangle anomaly. 
The strategy is closely parallel to the one presented in Ref.\cite{Son:2009tf}: one imposes the positivity condition
on the local entropy production, which turns out powerful enough to allow the determination of the transport coefficients.
Compared to the
original study in Ref.\cite{Son:2009tf} of the general non-conformal case, we will see that the derivation in the conformal case has some subtle differences, although the results will be identical. This first order case is a warm-up exercise prior to a more elaborate study of the second order
hydrodynamics in the next subsection. For simplicity we consider a single $U(1)$ current.

The basic Weyl covariant equations of hydrodynamics are
\bear
\CD_\mu T^{\mu\nu} &=& F^{\nu\alpha} j_\alpha\quad,\nonumber\\
\CD_\mu j^\mu&=& {\kappa\over 8}\,\epsilon^{\mu\nu\alpha\beta} F_{\mu\nu} F_{\alpha\beta} = \kappa\,E^\mu B_\mu\quad,\label{basic}
\eear
where $F_{\mu\nu}=\partial_\mu A_\nu-\partial_\nu A_\mu=\CD_\mu A_\nu-\CD_\nu A_\mu$ is the field strength tensor of external gauge potential $A_\mu$, which has a Weyl weight $w=0$, and we define
\bear
E^\mu=F^{\mu\nu}u_\nu\quad,\quad B^\mu={1\over 2}\epsilon^{\mu\nu\alpha\beta} u_\nu F_{\alpha\beta}\quad,
\eear
both of which have $w=+3$. They are also transverse, $E^\mu u_\mu=B^\mu u_\mu=0$. One can check that (\ref{basic}) including the anomaly is consistent with Weyl weights.
We then write the constitutive relations of $T^{\mu\nu}$ and $j^\mu$ in terms of the local thermodynamic variables,
\bear
T^{\mu\nu}&=& 4p u^\mu u^\nu + p g^{\mu\nu} +\tau_{(1)}^{\mu\nu}+\tau_{(2)}^{\mu\nu}+\cdots=
4p u^\mu u^\nu + p g^{\mu\nu} +\tau^{\mu\nu}\quad,\nonumber\\
j^\mu&=& n u^\mu +\nu^\mu_{(1)}+\nu^\mu_{(2)}+\cdots=n u^\mu +\nu^\mu\quad,\label{con}
\eear
where the subscripts in the viscous terms denote the number of derivatives each term contains. In the Landau frame defined as
\be
T^{\mu\nu}u_\nu = 3p u^\mu\quad,\quad j^\mu u_\mu = -n\quad,
\ee
the viscous terms must be strictly transverse order by order,
\be
\tau_{(n)}^{\mu\nu} u_\nu = \nu^\mu_{(n)} u_\mu=0\quad.
\ee
Following Ref.\cite{Son:2009tf} we consider $F_{\mu\nu}$ as of first order in derivative, which is a weak field limit.

Among the thermodynamic variables $(p,T,\mu,u^\mu)$ with $u^\mu u_\mu=-1$ or their combinations, there are five independent
ones that can be chosen to describe a local property of the plasma; for convenience, we choose two scalars $(T,\bar\mu\equiv {\mu\over T})$
and $u^\mu$. The number of equations in (\ref{basic}) is also five and the time evolution of the system 
is well-defined.  
In writing down possible viscous terms containing derivatives, one can use the basic equations of motion in (\ref{basic})
to replace certain first order derivative terms with higher order derivatives, effectively removing them at a given fixed
order in derivatives. Explicitly, the first equation in (\ref{basic}) with (\ref{con}) gives
\be
 4\left(\CD_\mu p\right) u^\mu u^\nu +\CD^\nu p +\CD_\mu \tau^{\mu\nu} = n E^\nu+F^{\nu\alpha}\nu_\alpha\quad.
\ee
Contracting with $u_\nu$, one obtains
\be
u^\mu \CD_\mu p = {1\over 3}\left(E^\mu \nu_\mu+u_\nu\CD_\mu \tau^{\mu\nu}\right)\quad,
\ee
and inserting this into the original equation, one gets
\be
\CD^\mu p = n E^\mu + F^{\mu\alpha}\nu_\alpha-{4\over 3} u^\mu\left(E^\alpha \nu_\alpha+u_\nu\CD_\alpha\tau^{\alpha\nu}\right)
-\CD_\nu\tau^{\mu\nu}\quad.\label{bas1}
\ee
Similarly, from the second equation in (\ref{basic}), one obtains
\be
u^\mu \CD_\mu n = -\CD_\mu \nu^\mu + \kappa\, E^\mu B_\mu\quad. \label{bas2}
\ee
Now, any scalar thermodynamic variable should be a function of two variables we choose, i.e. $(T,\bar\mu)$; note that
$T$ has the Weyl weight $w=+1$ and $\bar\mu$ has $w=0$. This means that a scalar $f$ with a weight $w$ can always be written as
\be
f=T^w \bar f=T^w \bar f(\bar\mu)\quad;
\ee
we will use this quite often.
Writing $p=T^4 \bar p$ and $n=T^3 \bar n$, one sees that eqns. (\ref{bas1}), (\ref{bas2}) can be used
to remove $\CD_\mu T$ and $u^\mu \CD_\mu \bar \mu$ in favor of $E_\mu$ and $\Delta_{\mu\nu}\CD^\nu\bar \mu$ up to
higher derivatives, where $\Delta_{\mu\nu}\equiv (g_{\mu\nu}+u_\mu u_\nu)$ is the projection operator to
the  space transverse to $u^\mu$. Therefore, one can always ignore them in writing down possible viscous terms 
at a given order in the derivative expansion. 
This means that the only derivative term of a scalar quantity one needs to consider in constructing the viscous terms is $\Delta_{\mu\nu}\CD^\nu\bar \mu$; indeed, 
any scalar is a function of $(T,\bar \mu)$ only, and the only other possible derivatives are $\CD_\mu u_\nu$.
This is a big simplification.
It will also appear useful  that $E_\mu$ can be replaced by
\be
E_\mu = {1\over n}\CD_\mu p  + {\rm higher\,\,derivatives}\quad.
\ee
Since the number of equations in (\ref{basic}) is five and we have used them to remove five components $\CD_\mu T$ and $u^\mu \CD_\mu\bar \mu$, no further reduction is possible from the equations of motion.

Following Ref.\cite{Son:2009tf}, one considers the local entropy production rate $\CD_\mu s^\mu=\nabla_\mu s^\mu$ where the equality is
valid since $s^\mu$ has $w=+4$, and imposes the positivity condition on all possible configurations, which will lead to
a few differential equations for the transport coefficients. It appears possible to solve them using conformal symmetry.
As usual, one starts from
\bear
u_\nu\CD_\mu T^{\mu\nu}+\mu \CD_\mu j^\mu&=& u_\nu F^{\mu\alpha}j_\alpha+\kappa\,\mu E^\mu B_\mu\nonumber\\
&=& u_\nu\left(4\left(\CD_\mu p\right) u^\mu u^\nu+\CD^\nu p +\CD_\mu\tau^{\mu\nu}\right)+\mu\left(\left(\CD_\mu n\right)u^\mu
+\CD_\mu \nu^\mu\right)\nonumber\\
&=& - u^\mu\CD_\mu(3p) + \mu u^\mu\CD_\mu n +u_\nu\CD_\mu \tau^{\mu\nu}+\mu \CD_\mu \nu^\mu\nonumber\\
&=& - u^\mu\CD_\mu \epsilon + \mu u^\mu\CD_\mu n -\left(\CD_\mu u_\nu\right) \tau^{\mu\nu}+\mu \CD_\mu \nu^\mu\quad,
\eear
using (\ref{prop}), $\epsilon=3p$ and $u_\nu \tau^{\mu\nu}=0$. From the thermodynamic relations ${\rm d}\epsilon=T{\rm d}s+\mu {\rm d}n$,
$\epsilon+p=4p=Ts+\mu n$, and the Weyl weights of each quantity, one can check that
\be
\CD_\mu\epsilon= T\CD_\mu s+\mu\CD_\mu n\quad;
\ee
and using this relation in the above formula after some algebra gives
\bear
T\CD_\mu\left(s u^\mu - \bar \mu \nu^\mu\right) = -\left(\CD_\mu u_\nu\right)\tau^{\mu\nu}-\left(
T\CD_\mu \bar\mu - E_\mu\right)\nu^\mu -C\,\mu E^\mu B_\mu\quad.
\eear
One proceeds by writing the entropy current $s^\mu$ 
in the derivative expansion as
\be
s^\mu = s u^\mu - \bar \mu \nu^\mu + s^\mu_{(1)}+s^\mu_{(2)}+\cdots\quad,
\ee
so that the total entropy production rate we want to keep positive definite is
\be
T\CD_\mu s^\mu = -\sigma_{\mu\nu}\tau^{\mu\nu}-\left(
T\CD_\mu \bar\mu - E_\mu\right)\nu^\mu -\kappa\,\mu E^\mu B_\mu+T\CD_\mu\left(s^\mu_{(1)}+s^\mu_{(2)}+\cdots\right)\ge 0\quad,\label{entropy}
\ee
where we define 
\be
\sigma_{\mu\nu}={1\over 2}\left(\CD_\mu u_\nu+\CD_\nu u_\mu\right)\quad,\quad
\omega_{\mu\nu}={1\over 2}\left(\CD_\mu u_\nu-\CD_\nu u_\mu\right)\quad,\quad \CD_\mu u_\nu=\sigma_{\mu\nu}+\omega_{\mu\nu}\quad.\label{sigome}
\ee
Note that $\sigma_{\mu\nu}$ and $\omega_{\mu\nu}$ have $w=-1$ and are already tansverse and traceless due to (\ref{prop}). The constraint 
(\ref{entropy}) is the main starting point in considering the entropy production; it can be rewritten in a simple form within the conformal formalism.
The task is to classify $\tau^{\mu\nu}$, $\nu^\mu$, and $s^\mu_{(n)}$ order by order in derivatives and to find
constraints on the transport coefficients associated to them by imposing positivity on the local entropy current (\ref{entropy}).

At first order in derivatives, independent available tensors are
\be
\Delta^{\mu\nu}\CD_\nu\bar\mu\quad,\quad \CD_\mu u_\nu=\sigma_{\mu\nu}+\omega_{\mu\nu}\quad,\quad E^\mu\quad;\quad
B^\mu\quad,\label{list1}
\ee
it is also useful to consider a transverse pseudo-vector of $w=+2$,
\be
\omega^\mu \equiv {1\over 2}\epsilon^{\mu\nu\alpha\beta}u_\nu\omega_{\alpha\beta}=
{1\over 2}\epsilon^{\mu\nu\alpha\beta}u_\nu\nabla_\alpha u_\beta\quad.
\ee
For the first order energy-momentum tensor $\tau_{(1)}^{\mu\nu}$ which must be transverse and traceless, there is only one
possibility as can be deduced from (\ref{list1}): $\sigma^{\mu\nu}$. Taking into account the positivity of entropy production (\ref{entropy}) it must assume the classic form
\be
\tau_{(1)}^{\mu\nu}=-2\eta \sigma^{\mu\nu}\quad(\eta>0)\quad,\label{em1}
\ee
where the shear viscosity $\eta$ has a weight $w=+3$.

For
$\nu^\mu_{(1)}$ and $s^\mu_{(1)}$, one needs to construct the transverse vectors and there are four of them: 
$\Delta^{\mu\nu}\CD_\nu\bar\mu$, $E^\mu$, $B^\mu$, and $\omega^\mu$. The positivity of (\ref{entropy}) allows to 
fix easily the dependence on the first two:
\bear
\nu^\mu_{(1)} &=& -\sigma\left(T\Delta^{\mu\nu}\CD_\nu\bar\mu-E^\mu\right) +\xi \omega^\mu +\xi_B B^\mu\quad,\\
s^\mu_{(1)} &=& D \omega^\mu+ D_B B^\mu \quad,\label{cur1}
\eear
with a positive conductivity $\sigma$ of weight $w=+1$; the remaining transport coefficients $\xi$ ($w=+2$), $\xi_B$ ($w=+1$), $D$ ($w=+2$), and $D_B$ ($w=+1$) are to be determined.
We will rigorously show in the next subsection using discrete symmetries that they should be proportional to the anomaly coefficient $\kappa$ as they originate from the anomaly, but for now we will simply let them be general possible terms as in Ref.\cite{Son:2009tf}; we then insert the above into (\ref{entropy}).

\vskip0.3cm
\noindent To proceed, we need the following two identities to be valid at all orders:
\be
\CD_\mu \omega^\mu =0\quad,\quad \CD_\mu B^\mu = -2 E^\mu\omega_\mu\quad.\label{iden1}
\ee
{\it Proof : } From (\ref{prop}) and (\ref{sigome}) one has
\bear
2 \CD_\mu \omega^\mu &=&  \epsilon^{\mu\nu\alpha\beta}\left(\CD_\mu u_\nu\right)\omega_{\alpha\beta}
+ \epsilon^{\mu\nu\alpha\beta}u_\nu \CD_\mu\CD_\alpha u_\beta\nonumber\\
&=& \epsilon^{\mu\nu\alpha\beta}\omega_{\mu\nu}\omega_{\alpha\beta}
+{1\over 2}\epsilon^{\mu\nu\alpha\beta}u_\nu [\CD_\mu,\CD_\alpha] u_\beta\quad.
\eear
Because $\omega_{\mu\nu}$ is transverse, that is, in the local rest frame of $u^i=0$ the only non-vanishing components are
$\omega_{ij}$ ($i,j=1,2,3$), the first term clearly vanishes. For the second piece, use the relation (\ref{rel2})
\be
[\CD_\mu,\CD_\alpha]u_\beta= -\CW_{\mu\alpha}u_\beta - \CR_{\mu\alpha\beta}^{\quad\,\,\delta}u_\delta\quad,
\ee
and a symmetry property \cite{Loganayagam:2008is}
\be
\CR_{\mu\alpha\beta}^{\quad\,\,\delta}+\CR_{\beta[\mu,\alpha]}^{\quad\,\,\,\,\,\,\delta}=0\quad,
\ee
so that
\be
2\CD_\mu\omega^\mu = 
-{1\over 2}\epsilon^{\mu\nu\alpha\beta}u_\nu \CR_{\mu\alpha\beta}^{\quad\,\,\delta}u_\delta
=-{1\over 6}\epsilon^{\mu\nu\alpha\beta}u_\nu\left( 
\CR_{\mu\alpha\beta}^{\quad\,\,\delta}+\CR_{\beta[\mu,\alpha]}^{\quad\,\,\,\,\,\,\delta}\right)u_\delta =0
\quad.
\ee
For the second identity, one has
\bear
2\CD_\mu B^\mu&=& \epsilon^{\mu\nu\alpha\beta}\left(\CD_\mu u_\nu\right)F_{\alpha\beta}+\epsilon^{\mu\nu\alpha\beta}
u_\nu\left(\CD_\mu F_{\alpha\beta}\right)\nonumber\\
&=& \epsilon^{\mu\nu\alpha\beta}\omega_{\mu\nu} F_{\alpha\beta}+\epsilon^{\mu\nu\alpha\beta}
u_\nu\left(\nabla_\mu F_{\alpha\beta}\right)\quad.
\eear
The second piece vanishes via Bianchi identity, while for the first piece it is most convenient to go to the local rest frame
where
\be
2\CD_\mu B^\mu= 2 \epsilon^{ijk0}\omega_{ij} F_{k0}\quad.
\ee
Noting that in this frame we have
\be
E_k=F_{k0}u^0\quad,\quad \omega^k={1\over 2}\epsilon^{k0ij}u_0 \omega_{ij}=-{1\over 2}u^0\epsilon^{ijk0}\omega_{ij}\quad,
\ee
one finally concludes that 
\be
2\CD_\mu B^\mu = -4 E_k \omega^k  = -4 E^\mu \omega_\mu \quad.\quad ({\rm QED})
\ee

\vskip0.3cm

Inserting (\ref{em1}), (\ref{cur1}) to (\ref{entropy}) and using (\ref{iden1}), one arrives at
\bear
T\CD_\mu s^\mu &=&2\eta \sigma_{\mu\nu}\sigma^{\mu\nu}+\sigma\left(T\Delta^{\mu\nu}\CD_\nu\bar\mu - E^\mu\right) 
\left(T\Delta_{\mu\alpha}\CD^\alpha \bar\mu - E_\mu\right)\nonumber\\
&-&\left(T\CD^\mu\bar\mu - E^\mu\right)\left(\xi \omega_\mu+\xi_B B_\mu\right)-\kappa\,\mu E^\mu B_\mu \nonumber\\
&+& T\left(\left(\CD_\mu D\right)\omega^\mu -2 D_B E_\mu\omega^\mu +\left(\CD_\mu D_B\right) B^\mu\right)\nonumber\\
&-& \sigma_{\mu\nu}\tau^{\mu\nu}_{(2)}- \left(T\CD_\mu\bar\mu - E_\mu\right)\nu^\mu_{(2)} + T\CD_\mu s^\mu_{(2)}+\cdots\quad,
\eear
where the first line is manifestly positive definite and the last line is of higher order. In considering arbitrary
configurations, it is important to remember that $E_\mu$ is not independent, but is equal to ${1\over n}\CD_\mu p$
up to higher derivative corrections. Writing $p=T^4\bar p(\bar\mu)$, this implies $E_\mu$ is given in terms of $\CD_\mu T$
and $\CD_\mu\bar\mu$. However, some other terms in the above, $\CD_\mu D$ and $\CD_\mu D_B$, are also
expressed in terms of $\CD_\mu T$ and $\CD_\mu\bar\mu$ as we write $D=T^2\bar D$ and $D_B=T \bar D_B$,
and therefore, $E_\mu$ is not completely independent of $\CD_\mu D$ and $\CD_\mu D_B$ as one considers arbitrary configurations
at the first order. The easiest systematic way to deal with this subtle difference from the non-conformal case is to simply replace $E_\mu$
with ${1\over n}\CD_\mu p$ by (\ref{bas1}):
\be
E^\mu= {1\over n} \CD^\mu p -{1\over n} F^{\mu\alpha}\nu_\alpha+{4\over 3n} u^\mu\left( E^\alpha \nu_\alpha
+u^\nu\CD_\alpha\tau^{\alpha\nu}\right) +{1\over n}\CD_\alpha \tau^{\alpha\mu}\quad,
\ee
at a given derivative order in considering arbitrary configurations. One then finds after some algebra
\bear
T\CD_\mu s^\mu &=&2\eta \sigma_{\mu\nu}\sigma^{\mu\nu}+\sigma\left(T\Delta^{\mu\nu}\CD_\nu\bar\mu - E^\mu\right) 
\left(T\Delta_{\mu\alpha}\CD^\alpha \bar\mu - E_\mu\right)\nonumber\\
&+&\left(-T\xi\CD_\mu\bar\mu +T\CD_\mu D+\left({\xi\over n} -{2T D_B\over n}\right) \CD_\mu p \right)\omega^\mu\nonumber\\
&+&\left( -T\xi_B \CD_\mu\bar\mu+T\CD_\mu D_B+\left({\xi_B\over n}-\kappa{\mu\over n}\right)\CD_\mu p        \right)B^\mu
\nonumber\\
&-&  \sigma_{\mu\nu}\tau^{\mu\nu}_{(2)}-\left(T\CD_\mu\bar\mu - E_\mu\right)\nu^\mu_{(2)} + T\CD_\mu s^\mu_{(2)}+\cdots \nonumber\\
&+&{1\over n}\left(-F^{\mu\alpha}\nu_\alpha+\CD_\alpha\tau^{\alpha\mu}\right)\left((\xi-2TD_B)\omega_\mu+(\xi_B-\kappa \mu)B_\mu\right)\quad;\label{start2}
\eear
the last line is a remnant from replacing $E_\mu$ by ${1\over n}\CD_\mu p$ using (\ref{bas1}), which is
in fact relevant when we discuss the second order viscous terms in the next subsection.

Because $\omega^\mu$ and $B^\mu$ can take arbitrary values and also can change their signs, the second and third lines
can easily overcome the first unless the coefficients vanish identically. This leads to two differential equations
\bear
-T\xi\CD_\mu\bar\mu +T\CD_\mu D+\left({\xi\over n} -{2T D_B\over n}\right) \CD_\mu p &=&0 \quad,\\
-T\xi_B \CD_\mu\bar\mu+T\CD_\mu D_B+\left({\xi_B\over n}-\kappa{\mu\over n}\right)\CD_\mu p&=& 0\quad.
\eear
Let us now substitute 
\be
p=T^4\bar p(\bar\mu)\quad,\quad (D,\xi)=T^2(\bar D,\bar\xi)\quad,\quad (D_B,\xi_B)=T(\bar D_B,\bar \xi_B)\quad,
\quad n=T^3 \bar n\quad,
\ee
upon which the above becomes
\bear
T^3\left(-\bar\xi+\bar D'+{1\over\bar n}\left(\bar\xi-2\bar D_B\right)\bar p' \right)\CD_\mu\bar\mu
+T^2 \left(2\bar D+{1\over \bar n}\left(\bar\xi-2\bar D_B\right) 4\bar p\right)\CD_\mu T &=&0\,,\nonumber\\
T^2\left(-\bar\xi_B+\bar D_B'+{1\over\bar n}\left(\bar\xi_B-\kappa\bar\mu \right)\bar p' \right)\CD_\mu\bar\mu
+T\left(\bar D_B+{1\over \bar n}\left(\bar\xi_B-\kappa \bar\mu\right) 4\bar p\right)\CD_\mu T &=&0\,,\nonumber
\eear
where prime denotes ${d\over d\bar\mu}$. As $T$ and $\bar\mu$ are independent, the coefficients in front of $\CD_\mu T$
and $\CD_\mu\bar\mu$ must vanish separately, so that one in fact gets four equations to solve.

\vskip0.3cm

\noindent We now need to prove the following relation:
\be
\bar p' = \bar n\quad.\label{pn}
\ee
{\it Proof : } Let us start from the basic thermodynamic relations 
$\epsilon+p=4p=Ts+\mu n$ and ${\rm d} \epsilon= 3{\rm d}p= T{\rm d}s+\mu{\rm d}n$, which give
\bear
4\bar p&=& \bar s+\bar\mu\bar n\quad,\\
 3\left(4T^3\bar p {\rm d}T+T^4 \bar p' {\rm d}\bar\mu\right)&=&
T\left(3T^2 \bar s {\rm d}T+T^3 \bar s'{\rm d}\bar\mu\right)+T\bar\mu\left(3T^2\bar n {\rm d}T+T^3 \bar n' {\rm d}\bar\mu\right)\quad.\nonumber
\eear  
As $T$ and $\bar\mu$ are independent variables, the coefficients of ${\rm d}T$ and ${\rm d}\bar\mu$
on both sides in the second relation should agree separately. The equation from ${\rm d}T$ is precisely the first relation,
while from ${\rm d}\bar\mu$ one obtains
\be
3\bar p'= \bar s'+\bar\mu \bar n'= 4\bar p'-(\bar\mu\bar n)' +\bar\mu\bar n'=4\bar p'-\bar n\quad,
\ee
where we use the first relation in the second equality. One gets $\bar p'=\bar n$ from the final expression. (QED)

\vskip0.3cm

Using (\ref{pn}), the four equations simplify as
\bear
\bar D'+2\bar D_B=0\quad&,&\quad \bar D_B' -\kappa\bar\mu =0\quad,\\
\bar\xi=2\bar D_B - {\bar n\over 2\bar p}\bar D\quad&,&\quad \bar\xi_B=\kappa \bar\mu - {\bar n\over 4\bar p}\bar D_B\quad,
\eear  
from which it is easy to integrate $\bar D$ and $\bar D_B$ as \footnote{We are neglecting possible
integration constants \cite{Neiman:2010zi}, which might be related to gravitational anomalies \cite{Landsteiner:2011cp}.}  
\be
\bar D={1\over 3}\kappa\bar\mu^3\quad,\quad \bar D_B= {1\over 2}\kappa \bar\mu^2\quad,\label{firres1}
\ee
and the second line finally gives
\be
\bar\xi = \kappa\left(\bar\mu^2 - {2\over 3}{\bar n\over 4\bar p} \bar\mu^3\right)\quad,\quad
\bar\xi_B=\kappa\left(\bar\mu-{1\over 2}{\bar n\over 4\bar p} \bar\mu^2\right)\quad.\label{firres2}
\ee
One can check that our results agree precisely with the general non-conformal results in Ref.\cite{Son:2009tf} upon using the conformal relation $\epsilon+p=4p$.

\subsection{Second order conformal hydrodynamics with anomaly}

In this subsection, we address our main objective of studying the second order viscous corrections to the energy-momentum
tensor and the current, with particular attention to the anomaly--induced effects.
Our starting point is Eq.(\ref{start2}) from the previous subsection: 
\bear
T\CD_\mu s^\mu &=&2\eta \sigma_{\mu\nu}\sigma^{\mu\nu}+\sigma\left(T\Delta^{\mu\nu}\CD_\nu\bar\mu - E^\mu\right) 
\left(T\Delta_{\mu\alpha}\CD^\alpha \bar\mu - E_\mu\right)\nonumber\\
&+&\left(-T\xi\CD_\mu\bar\mu +T\CD_\mu D+\left({\xi\over n} -{2T D_B\over n}\right) \CD_\mu p \right)\omega^\mu\nonumber\\
&+&\left( -T\xi_B \CD_\mu\bar\mu+T\CD_\mu D_B+\left({\xi_B\over n}-\kappa{\mu\over n}\right)\CD_\mu p        \right)B^\mu
\nonumber\\
&-& \sigma_{\mu\nu}\tau^{\mu\nu}_{(2)}-\left(T\CD_\mu\bar\mu - E_\mu\right)\nu^\mu_{(2)} + T\CD_\mu s^\mu_{(2)}+\cdots \nonumber\\
&+&{1\over n}\left(-F^{\mu\alpha}\nu_{\alpha(1)}+\CD_\alpha\tau^{\alpha\mu}_{(1)}\right)\left((\xi-2TD_B)\omega_\mu+(\xi_B-\kappa \mu)B_\mu\right)\quad.\label{start22}
\eear
The first three lines are already taken care of in the previous subsection; for example the second and the third lines simply vanish when we use our solutions for the first order transport coefficients $\xi,\xi_B,D,D_B$. The last two lines
are what is important in this subsection.
We are interested in possible second order viscous corrections $\tau^{\mu\nu}_{(2)}$, $\nu^\mu_{(2)}$, $s^\mu_{(2)}$, and
by considering the entropy production we would like to constrain the transport coefficients associated with them as much as possible. Although we will classify all possible independent second order viscous corrections, we only focus on the 
transport coefficients that necessarily arise from the triangle anomaly in our consideration of entropy production. In other words,
we are going to specify to a subclass of possible terms with transport coefficients that are linear in the anomaly constant $\kappa$.
Due to a selection rule from discrete symmetries of charge conjugation $C$ and parity $P$ that we will discuss shortly,
there is no interference in (\ref{start22}) between the anomaly-induced terms that we focus on and other ``non-anomalous'' terms, so that
one can meaningfully separate them in the consideration of entropy production.
We leave for the future the task of fully exploring the constraints from entropy production on all possible ``non-anomalous" transport coefficients that 
we list. 

Let us begin with $\nu^\mu_{(2)}$ and $s^\mu_{(2)}$ and write down all possible independent Weyl covariant transverse vectors
that include two derivatives. As discussed before, hydrodynamic equations of motion can be used to remove
covariant derivatives of scalars such as $\CD_\mu T$ in favor of $E^\mu$ and $\Delta^{\mu\nu}\CD_\nu\bar\mu$. 
Other possible tensors of use are $\CD_\mu u_\nu$ and $B^\mu$. It is quite a tedious job to list
all possible combinations one can construct and to identify independent components, so we simply present the resulting fifteen terms:
\bear
&&\sigma^{\mu\nu}\CD_\nu\bar\mu\,\,,\,\,\omega^{\mu\nu}\CD_\nu\bar\mu\,\,,\,\,
\Delta^{\mu\nu}\CD^\alpha \sigma_{\nu\alpha}\,\,,\,\,\Delta^{\mu\nu}\CD^\alpha \omega_{\nu\alpha}
\,\,,\,\,\sigma^{\mu\nu}\omega_\nu\,\,,\nonumber\\
&& \sigma^{\mu\nu}E_\nu\,\,,\,\,\sigma^{\mu\nu}B_\nu\,\,,\,\,\omega^{\mu\nu}E_\nu\,\,,\,\,\omega^{\mu\nu}B_\nu\,\,,u^\nu\CD_\nu E^\mu\,\,,\\
&& \epsilon^{\mu\nu\alpha\beta}u_\nu E_\alpha\CD_\beta\bar\mu\,\,,\,\,
\epsilon^{\mu\nu\alpha\beta}u_\nu B_\alpha\CD_\beta\bar\mu\,\,,\,\,\epsilon^{\mu\nu\alpha\beta}u_\nu E_\alpha B_\beta\,\,,\,\,
\epsilon^{\mu\nu\alpha\beta}u_\nu \CD_\alpha E_\beta\,\,,\,\,\epsilon^{\mu\nu\alpha\beta}u_\nu \CD_\alpha B_\beta\,\,.\nonumber
\eear 
The five terms in the first line were found before in Refs.\cite{Erdmenger:2008rm,Banerjee:2008th}, and the rest are new. There are a few details involved in showing that these are indeed 
all possible independent terms. For example, $u^\nu\CD_\nu B^\mu$ is not included in the above because it is related to
the term $\epsilon^{\mu\nu\alpha\beta}u_\nu \CD_\alpha E_\beta$ by Bianchi identity. Another possibility is to use the second order
Weyl curvature tensors to construct terms such as $\CW^{\mu\nu}u_\nu$ and $\Delta^{\mu\nu}\CR_{\nu\alpha}u^\alpha$, but one can show that
\be
\CD^\nu\CD_\mu u_\nu = \CD^\nu\left(\sigma_{\mu\nu}+\omega_{\mu\nu}\right)=-\CW_{\mu\nu}u^\nu-\CR_{\mu\nu}u^\nu\quad,
\ee
and moreover, using the relation 
\be
\CW_{\mu\nu}={1\over 4p}[\CD_\mu,\CD_\nu]p={1\over 4p}\left(\CD_\mu\left(n E_\nu\right)-\CD_\nu\left(n E_\mu\right)\right)+{\rm higher\,\,orders}\quad,
\ee
one can easily check that $\CW^{\mu\nu}u_\nu$ and $\Delta^{\mu\nu}\CR_{\nu\alpha}u^\alpha$ are already included in the
above list. We leave other checks to readers.

As we are interested in terms that are necessarily linear in the anomaly coefficient $\kappa$, we need a systematic method to identify
such terms. We will use two discrete symmetries; charge conjugation $C$ and space parity in the local rest frame $P$.
Consider the basic hydrodynamic equation
\be
\CD_\mu j^\mu = \kappa E^\mu B_\mu\quad.
\ee
Under $(C,P)$, the spatial component of $j^\mu$ is $(-1,-1)$. Similarly, $E^\mu$ has $(-1,-1)$ and $B^\mu$ has $(-1,+1)$ \footnote{We define our $P$ such that the gauge potential $A_\mu$ transforms as $A_0\to A_0$ and $A^i\to-A^i$. Alternatively,
one can introduce an additional overall negative sign as in the case of axial symmetry in QCD. It is a matter of interchanging $P\leftrightarrow CP$, and as long as one adheres to a chosen definition, it would not make a difference for our purpose. }.
The covariant derivative $\CD_\mu$ has $(+1,-1)$. This tells us that $\kappa$ has $(C,P)=(-1,-1)$.
From the constitutive relation $j^\mu=n u^\mu+\cdots$, the charge density $n$ and the chemical potential $\bar\mu$
have $(C,P)=(-1,+1)$. The combination $(\kappa\bar\mu)$ then has $(C,P)=(+1,-1)$. Naturally, the entropy current $s^\mu$
has $(C,P)=(+1,-1)$.
The usefulness of these discrete charges is exposed by the following observation: when we write $\nu^\mu_{(2)}$ or $s^\mu_{(2)}$
as a linear combination of the terms in the above list, the transport coefficient in front of each term should have
a well-defined $(C,P)$ that is easily derived by comparing the $(C,P)$ of $\nu^\mu_{(2)}$ or $s^\mu_{(2)}$ with that of
each term in the list. These transport coefficients are conformal scalars of some weight $w$, so that they can be generically
written as $T^w f(\bar\mu,\kappa)$. Since they have definite $(C,P)$ and $(\bar\mu,\kappa, \kappa\bar\mu)$ all have 
different $(C,P)$'s, there are only four possibilities of the form of $f(\bar\mu,\kappa)$;
\bear
(C,P)=(+1,+1)\quad&:&\quad f(\bar\mu,\kappa)= g(\bar\mu^2,\kappa^2)\nonumber\\
(C,P)=(-1,+1)\quad&:&\quad f(\bar\mu,\kappa)=\bar\mu g(\bar\mu^2,\kappa^2)\nonumber\\
(C,P)=(-1,-1)\quad&:&\quad f(\bar\mu,\kappa)= \kappa g(\bar\mu^2,\kappa^2)\nonumber\\
(C,P)=(+1,-1)\quad&:&\quad f(\bar\mu,\kappa)= \kappa\bar\mu g(\bar\mu^2,\kappa^2)\nonumber
\eear
Therefore one can systematically select the terms that necessarily come from the anomaly by
choosing only the terms whose transport coefficients have $(C,P)=(\pm 1,-1)$.
In retrospect, the first order transport coefficients $\xi,\xi_B,D,D_B$ have precisely such $(C,P)$ charges.

With the help of this criterion, we find that five terms in the above list can be identified as originating from the anomaly:
\be
\sigma^{\mu\nu}\omega_\nu\quad,\quad \sigma^{\mu\nu}B_\nu\quad,\quad \omega^{\mu\nu}B_\nu\quad,\quad
\epsilon^{\mu\nu\alpha\beta}u_\nu E_\alpha \CD_\beta\bar\mu\quad,\quad 
\epsilon^{\mu\nu\alpha\beta}u_\nu \CD_\alpha E_\beta\quad,
\ee
and we introduce ten transport coefficients associated with them as
\bear
\nu^\mu_{(2){\rm anomaly}}&=& \xi_1 
\sigma^{\mu\nu}\omega_\nu+\xi_2 \sigma^{\mu\nu}B_\nu+\xi_3 \omega^{\mu\nu}B_\nu+\xi_4
\epsilon^{\mu\nu\alpha\beta}u_\nu E_\alpha \CD_\beta\bar\mu+\xi_5 
\epsilon^{\mu\nu\alpha\beta}u_\nu \CD_\alpha E_\beta\,,\nonumber\\
s^\mu_{(2){\rm anomaly}}&=& D_1 
\sigma^{\mu\nu}\omega_\nu+D_2 \sigma^{\mu\nu}B_\nu+D_3 \omega^{\mu\nu}B_\nu+
D_4
\epsilon^{\mu\nu\alpha\beta}u_\nu E_\alpha \CD_\beta\bar\mu+D_5 
\epsilon^{\mu\nu\alpha\beta}u_\nu \CD_\alpha E_\beta\,.\nonumber\\\label{trans1}
\eear

We perform a similar procedure for $\tau^{\mu\nu}_{(2)}$ which has a somewhat more complicated structure.
Refs.\cite{Erdmenger:2008rm,Banerjee:2008th} listed all possible independent terms without including external electric/magnetic fields, and 
it is not difficult to extend their results including external electromagnetic fields.  
Defining the projection operator to transverse, traceless and symmetric components as in Ref.\cite{Banerjee:2008th},
\be
\Pi^{\alpha\beta}_{\mu\nu} = {1\over 2}\left(\Delta_\mu^\alpha\Delta_\nu^\beta+\Delta_\mu^\beta\Delta_\nu^\alpha -{2\over 3}
\Delta^{\alpha\beta}\Delta_{\mu\nu}\right)\quad,
\ee
the independent possible terms are
\bear
&&u^\alpha\CD_\alpha\sigma^{\mu\nu}\,\,,\,\,\Pi^{\mu\nu}_{\alpha\beta}\sigma^{\alpha}_{\,\,\,\gamma}\sigma^{\gamma\beta}\,\,,\,\,\Pi^{\mu\nu}_{\alpha\beta}\sigma^{\alpha}_{\,\,\,\gamma}\omega^{\gamma\beta}\,\,,\,\,
\Pi^{\mu\nu}_{\alpha\beta}\omega^{\alpha}_{\,\,\,\gamma}\omega^{\gamma\beta}
\,\,,\,\,\Pi^{\mu\nu}_{\alpha\beta}\CD^\alpha\omega^\beta\,\,,\,\,
\Pi^{\mu\nu}_{\alpha\beta}\CD^\alpha\CD^\beta\bar\mu\,\,,\nonumber\\
&&\Pi^{\mu\nu}_{\alpha\beta}\CD^\alpha\bar\mu\CD^\beta\bar\mu\,\,,\,\,\Pi^{\mu\nu}_{\alpha\beta}\omega^\alpha\CD^\beta\bar\mu\,\,,\,\,\Pi^{\mu\nu}_{\alpha\beta}\epsilon^{\gamma\delta\eta\alpha}\sigma^{\beta}_{\,\,\,\eta}
u_\gamma\CD_\delta\bar\mu\,\,,\,\,\Pi^{\mu\nu}_{\alpha\beta}\CD^\alpha E^\beta\,\,,\,\,
\Pi^{\mu\nu}_{\alpha\beta}\CD^\alpha B^\beta\,\,,\,\,\Pi^{\mu\nu}_{\alpha\beta}E^\alpha\CD^\beta\bar\mu\,\,,\nonumber\\
&& \Pi^{\mu\nu}_{\alpha\beta}B^\alpha\CD^\beta\bar\mu\,\,,\,\,\Pi^{\mu\nu}_{\alpha\beta}E^\alpha E^\beta\,\,,\,\,\Pi^{\mu\nu}_{\alpha\beta}E^\alpha B^\beta\,\,,\,\,\Pi^{\mu\nu}_{\alpha\beta}B^\alpha B^\beta
\,\,,\,\,\Pi^{\mu\nu}_{\alpha\beta}\epsilon^{\gamma\delta\eta\alpha}\sigma^{\beta}_{\,\,\,\eta}
u_\gamma E_\delta\,\,,\,\,\Pi^{\mu\nu}_{\alpha\beta}\epsilon^{\gamma\delta\eta\alpha}\sigma^{\beta}_{\,\,\,\eta}
u_\gamma B_\delta\,\,,\nonumber\\
&&\Pi^{\mu\nu}_{\alpha\beta}\omega^\alpha E^\beta\,\,,\,\,
\Pi^{\mu\nu}_{\alpha\beta}\omega^\alpha B^\beta\,\,,\,\,C^{\mu\alpha\nu\beta}u_\alpha u_\beta\,\,,\,\,
\epsilon^{\mu\alpha\beta\gamma}\epsilon^{\nu\delta\eta\lambda}C_{\alpha\beta\delta\eta}u_\gamma u_\lambda\,\,,\,\,
\Pi^{\mu\nu}_{\alpha\beta}\epsilon^{\alpha\gamma\delta\eta}C_{\gamma\delta}^{\,\,\,\,\,\,\beta\lambda}u_\eta u_\lambda\,,\nonumber
\eear
where $C_{\mu\nu\alpha\beta}$ is the conformal Weyl tensor of the background metric.
Using the $(C,P)$ symmetries, one can pick up eight terms that should be linear in the anomaly $\kappa$,
and we introduce transport coefficients for them as
\bear
\tau^{\mu\nu}_{(2){\rm anomaly}}&=&
\lambda_1 \Pi^{\mu\nu}_{\alpha\beta}\CD^\alpha\omega^\beta+\lambda_2 \Pi^{\mu\nu}_{\alpha\beta}\omega^\alpha\CD^\beta\bar\mu+\lambda_3
\Pi^{\mu\nu}_{\alpha\beta}\epsilon^{\gamma\delta\eta\alpha}\sigma^{\beta}_{\,\,\,\eta}
u_\gamma\CD_\delta\bar\mu+\lambda_4\Pi^{\mu\nu}_{\alpha\beta}\CD^\alpha B^\beta\nonumber\\
&+&\lambda_5 \Pi^{\mu\nu}_{\alpha\beta}B^\alpha\CD^\beta\bar\mu+\lambda_6
\Pi^{\mu\nu}_{\alpha\beta}E^\alpha B^\beta+\lambda_7 
\Pi^{\mu\nu}_{\alpha\beta}\epsilon^{\gamma\delta\eta\alpha}\sigma^{\beta}_{\,\,\,\eta}
u_\gamma E_\delta+\lambda_8 \Pi^{\mu\nu}_{\alpha\beta}\omega^\alpha E^\beta\,.\nonumber\\
\label{trans2}
\eear
The eighteen transport coefficients $\xi_i,D_i,\lambda_j$ ($i=1,\cdots, 5$, $j=1,\cdots, 8$)
in (\ref{trans1}) and (\ref{trans2}) are the most general second order transport coefficients
of a conformal plasma that originate from the underlying triangle anomaly.

\subsection{Time reversal invariance, anomaly and entropy production}\label{time-reverse}

Let us now motivate the main guiding principle that we propose and are going to use extensively throughout this paper -- namely, that the anomaly-induced ${\cal T}$-even terms should not contribute to the entropy production. To illustrate the significance of discrete symmetries, let us consider first the behavior of contributions to the entropy under  the spatial parity. The anomalous contributions to 
the entropy production are special in that
they change sign under spatial parity transformation $P$. Suppose there were a contribution to the entropy production from the anomalous terms we identify; then in the parity-flipped mirror world, this contribution
would become negative. Thinking of entropy production as originating from some dissipative work,
this is very unnatural. This consideration gives us the first hint that the anomalous terms should not contribute to the entropy production. 

\vskip0.3cm

The vanishing of the entropy production from the anomaly-induced terms has a simple physical meaning -- the corresponding {\it anomalous currents are non-dissipative}.
This rather unusual property originates from the fact that the  anomalous current is associated with the zero fermion modes, and the number of these zero modes is related to the topology of gauge fields by the Atiyah-Singer index theorem. Since the topology of gauge fields is determined at the boundary of the fluid, the processes in the bulk cannot lead to the dissipation of anomalous currents. This consideration can be 
made more formal by considering yet another discrete symmetry of the transport coefficients - time reversal ${\cal T}$. The ``usual" electric conductivity $\sigma$ is a ${\cal T}$-odd quantity, as can be easily seen from Ohm's law $J^i = \sigma E^i$: the electric field is ${\cal T}$-even, whereas the electric current  $J^i$ is ${\cal T}$-odd. On the other hand, let us consider the quantum Hall effect as an example of anomalous current in $(2+1)$ dimensions. The quantum Hall conductance is a ${\cal T}$-even quantity, as it is associated with a ${\cal T}$-odd magnetic field. 
The corresponding Hall current is non-dissipative, and the conductance of the integer quantum Hall effect is given by the Chern numbers of vector bundles associated with the energy bands of the Hamiltonian operator \cite{Thouless:1982zz}. In physical terms, the dissipative transport coefficients are described in terms of the response of the states near the Fermi energy, whereas the non-dissipative ones involve all of the states below the Fermi energy. The anomalous chiral magnetic current can be thought of as a quantum phenomenon that involves the entire Dirac sea \cite{Kharzeev:2010ym} (reflecting Gribov's view  of ``anomalies as a manifestation of the high momentum collective motion in the vacuum" \cite{Gribov:1981ku,Dokshitzer:2004ie}), and it is thus natural to expect that it is of non-dissipative, reversible nature. Indeed, the chiral magnetic conductivity $\sigma_\chi$ \cite{Kharzeev:2009pj} defined by $\vec{J} = \sigma_\chi \vec{B}$ is a manifestly ${\cal T}$-even quantity as it relates  magnetic field and electric current both of which are ${\cal T}$-odd. We note that this feature of anomalous currents makes them 
potentially important in various applications that include quantum computing, see e.g. \cite{Murakami}.

\vskip0.3cm

We thus conjecture that the terms originating from the anomaly, {\it i.e.} the terms
that are linear in $\kappa$, do not contribute to the net entropy production at all orders.
The first order result in the previous subsection obeys this principle -- the first order contribution coming from
anomaly vanishes identically. The validity of this claim in the present second order will be 
evidenced shortly by a non-trivial test of our results  against  the 
existing holographic computations of some of our transport coefficients. 
  
\subsection{Constraints from time reversal invariance}\label{hol_check}
  
  Let us now impose the constraints of time reversal invariance (no entropy production from the anomaly) on the transport coefficients.
We are interested in only the last two lines in (\ref{start22}) because the previous lines are already taken
care of at pevious orders. The last line is an important remnant from the first order computation, and
we can insert $\nu^\mu_{(1)}=-\sigma\left(T\Delta^{\mu\nu}\CD_\nu\bar\mu - E^\mu\right)$ and $\tau^{\mu\nu}_{(1)}=-2\eta\sigma^{\mu\nu}$ to get the terms that are linear in anomaly coefficient.
After some manipulations that use the local rest frame expressions
\be
F^{ij}B_i =0\quad,\quad F^{ij}\omega_i=\epsilon^{j\nu\alpha\beta}u_\nu B_\alpha\omega_\beta=\omega^{j\nu}B_\nu\quad,
\ee
the last line in (\ref{start22}) can be written as
\be
{\sigma\over n}\left(\xi-2T D_B\right)\left(T\Delta_{\mu\alpha}\CD^\alpha\bar\mu - E_\mu\right)\omega^{\mu\nu}B_\nu -{2\over n}\CD_\mu\left(\eta \sigma^{\mu\nu}\right)
\left(\left(\xi-2TD_B\right)\omega_\nu+\left(\xi_B-\kappa\mu\right)B_\nu\right)\quad.
\ee
The most complicated part is to compute $\CD_\mu s^\mu_{(2)}$ and to gather the independent components.
There are a few non-trivial steps that we have to describe:
\bear
\CD_\mu\left(\epsilon^{\mu\nu\alpha\beta}u_\nu E_\alpha \CD_\beta\bar\mu\right)
=\epsilon^{\mu\nu\alpha\beta}\omega_{\mu\nu} E_\alpha\CD_\beta\bar\mu
+\epsilon^{\mu\nu\alpha\beta}u_\nu\left(\CD_\mu E_\alpha\right)\CD_\beta\bar\mu\quad,
\eear
where we used $[\CD_\mu,\CD_\beta]\bar\mu =0$ because $\bar\mu$ has $w=0$. In the first term,
both $\omega_{\mu\nu}$ and $E_\alpha$ are transverse so that $\CD_\beta\bar\mu$ should necessarily be
$u^\mu \CD_\mu\bar\mu$ which can be removed by using equations of motion as we pointed out before;
therefore, only the last piece survives. Another term is
\bear
\CD_\mu\left(\epsilon^{\mu\nu\alpha\beta}u_\nu \CD_\alpha E_\beta\right)
&=&\epsilon^{\mu\nu\alpha\beta}\omega_{\mu\nu}\CD_\alpha E_\beta
+{1\over 2}\epsilon^{\mu\nu\alpha\beta} u_\nu [\CD_\mu,\CD_\alpha]E_\beta\nonumber\\
&=&\epsilon^{\mu\nu\alpha\beta}\omega_{\mu\nu}\CD_\alpha E_\beta
+{1\over 2}\epsilon^{\mu\nu\alpha\beta} u_\nu \left(\CW_{\mu\alpha}E_\beta -\CR_{\mu\alpha\beta}^{\,\,\,\,\,\,\,\,\,\,\gamma} E_\gamma\right)\quad,
\eear
and using symmetry of $\CR_{\mu\alpha\beta}^{\,\,\,\,\,\,\,\,\,\,\gamma}+\CR_{\beta[\mu,\alpha]}^{\,\,\,\,\,\,\,\,\,\,\gamma}=0$,
the last term is equal to zero. Also, from $\CW_{\mu\alpha}={1\over 4p}[\CD_\mu,\CD_\alpha]p$
and $\CD_\mu p \approx n E_\mu$ up to higher order derivatives, one can easily show that the final result is
\be
\CD_\mu\left(\epsilon^{\mu\nu\alpha\beta}u_\nu \CD_\alpha E_\beta\right)=
\epsilon^{\mu\nu\alpha\beta}\omega_{\mu\nu}\CD_\alpha E_\beta
+{n\over 4p}\epsilon^{\mu\nu\alpha\beta} u_\nu\left(\CD_{\mu}E_{\alpha}\right) E_\beta +{\rm higher\,derivatives}\quad.
\ee

Using the above, the second order entropy production coming from the last two lines in (\ref{start22}) becomes
\bear
&-\sigma_{\mu\nu}\Bigg(
\lambda_1 \Pi^{\mu\nu}_{\alpha\beta}\CD^\alpha\omega^\beta+\lambda_2 \Pi^{\mu\nu}_{\alpha\beta}\omega^\alpha\CD^\beta\bar\mu+\lambda_4\Pi^{\mu\nu}_{\alpha\beta}\CD^\alpha B^\beta
&\nonumber\\&
+\lambda_5 \Pi^{\mu\nu}_{\alpha\beta}B^\alpha\CD^\beta\bar\mu+\lambda_6
\Pi^{\mu\nu}_{\alpha\beta}E^\alpha B^\beta+\lambda_8 \Pi^{\mu\nu}_{\alpha\beta}\omega^\alpha E^\beta\Bigg)&\nonumber\\&
-\left(T\Delta_{\mu\alpha}\CD^\alpha\bar\mu - E_\mu\right)
\Bigg(\xi_1 
\sigma^{\mu\nu}\omega_\nu+\xi_2 \sigma^{\mu\nu}B_\nu+\xi_3 \omega^{\mu\nu}B_\nu+\xi_5 
\epsilon^{\mu\nu\alpha\beta}u_\nu \CD_\alpha E_\beta\Bigg)&\nonumber\\
&+T\Bigg(
\left(\CD_\mu D_1\right) 
\sigma^{\mu\nu}\omega_\nu+\left(\CD_\mu D_2\right) \sigma^{\mu\nu}B_\nu+\left(\CD_\mu D_3\right) \omega^{\mu\nu}B_\nu
&\nonumber\\&
+\left(\CD_\mu D_4\right)
\epsilon^{\mu\nu\alpha\beta}u_\nu E_\alpha \CD_\beta\bar\mu+\left(\CD_\mu D_5\right) 
\epsilon^{\mu\nu\alpha\beta}u_\nu \CD_\alpha E_\beta\Bigg)&\nonumber\\
&+T\Bigg(D_1\left(\left(\CD_\mu\sigma^{\mu\nu}\right)\omega_\nu+\sigma^{\mu\nu}\CD_\mu \omega_\nu\right)
+D_2\left(\left(\CD_\mu\sigma^{\mu\nu}\right) B_\nu +\sigma^{\mu\nu}\CD_\mu B_\nu\right)&\nonumber\\
&+D_3\left(\left(\CD_\mu\omega^{\mu\nu}\right) B_\nu +\omega^{\mu\nu}\CD_\mu B_\nu\right)
+D_4\epsilon^{\mu\nu\alpha\beta} u_\nu\left(\CD_\mu E_\alpha\right)\CD_\beta\bar\mu
&\nonumber\\
&
+D_5\left(
\epsilon^{\mu\nu\alpha\beta}\omega_{\mu\nu}\CD_\alpha E_\beta
+{n\over 4p}\epsilon^{\mu\nu\alpha\beta} u_\nu\left(\CD_{\mu}E_{\alpha}\right) E_\beta\right)\Bigg)&\nonumber\\
&+{\sigma\over n}\left(\xi-2T D_B\right)\left(T\Delta_{\mu\alpha}\CD^\alpha\bar\mu - E_\mu\right)\omega^{\mu\nu}B_\nu -{2\over n}\CD_\mu\left(\eta \sigma^{\mu\nu}\right)
\left(\left(\xi-2TD_B\right)\omega_\nu+\left(\xi_B-\kappa\mu\right)B_\nu\right)&\,.\nonumber\\
\eear
Note that the terms with $\lambda_3$, $\lambda_7$ and $\xi_4$ disappear identically, so
these transport coefficients are simply unconstrained by our method.
As before, one should replace $E_\mu$ by ${1\over n}\CD_\mu p$ in the above, and impose the condition that
the total coefficient of each independent component should vanish. We list each component and the equation imposed by the condition of zero entropy production as follows:
\begin{itemize}
 \item From the component $\sigma^{\mu\nu}\omega_\nu$, one obtains
\bear
-\lambda_2 \CD_\mu\bar\mu-{\lambda_8\over n}\CD_\mu p -\xi_1\left(T\CD_\mu\bar\mu -{1\over n}\CD_\mu p\right)
+T\CD_\mu D_1 -{2\over n}\left(\xi-2T D_B\right)\CD_\mu \eta =0\,.\nonumber
\eear
The transport coefficients are all conformal scalars of some weight, and one can generally write them as
$f=T^w \bar f(\bar\mu)$. Inserting this form and noting that $\CD_\mu T$ and $\CD_\mu\bar\mu$ are independent,
the coefficients in front of these should vanish separately, so that the above equation in fact provides two equations. Writing $\lambda_2=T^2\bar\lambda_2$, $\lambda_8=T \bar\lambda_8$, $\xi_1=T\bar\xi_1$, $D_1=T\bar D_1$, $\eta=T^3\bar\eta$ according to their conformal weights, and using (\ref{pn}) $\bar p'=\bar n$, one gets
two equations as
\bear
-\bar\lambda_2-\bar\lambda_8 +\bar D_1' -{2\over\bar n}\left(\bar\xi-2\bar D_B\right)\bar\eta'&=&0\quad,\label{en1}\\
-4\bar\lambda_8{\bar p\over\bar n}+4\bar\xi_1 {\bar p\over\bar n}+\bar D_1 -{6\over \bar n}\left(\bar \xi 
-2\bar D_B\right)\bar\eta &=&0 \quad,\label{en2}
\eear
where prime as usual denotes ${d\over d\bar\mu}$.

\item From $\left(\CD_\mu \sigma^{\mu\nu}\right)\omega_\nu$, one gets
\be
\bar D_1 - {2\bar\eta\over \bar n}\left(\bar\xi-2\bar D_B\right) =0\quad. \label{en3}
\ee

\item From $\left(\CD_\mu \sigma^{\mu\nu}\right)B_\nu$, one obtains
\be
\bar D_2 - {2\bar\eta\over \bar n}\left(\bar\xi_B-\kappa\bar\mu\right) =0\quad,\label{en4}
\ee
where $D_2=\bar D_2$ due to its zero conformal weight.

\item From $\sigma^{\mu\nu}B_\nu$, one has
\bear
-\lambda_5\CD_\mu\bar\mu-{\lambda_6\over n}\CD_\mu p -\xi_2\left(T\CD_\mu\bar\mu-{1\over n}\CD_\mu p\right)
+T\CD_\mu D_2 -{2\over n}\left(\xi_B-\kappa\mu\right)\CD_\mu\eta =0\,,\nonumber
\eear
which leads to two equations upon writing $\lambda_5=T\bar\lambda_5$, $\lambda_6=\bar\lambda_6$, $\xi_2=\bar\xi_2$,
\bear
-\bar\lambda_5 -\bar\lambda_6 +\bar D_2'-{2\over\bar n}\left(\bar\xi_B-\kappa\bar\mu\right)\bar\eta' =0\quad,
\label{en5}\\
-4\bar\lambda_6 {\bar p\over\bar n}+4\bar \xi_2 {\bar p\over\bar n}-{6\over\bar n}\left(\bar\xi_B-\kappa\bar\mu\right) \bar\eta =0\quad.\label{en6}
\eear

\item From $\left(\CD_\mu\omega^{\mu\nu}\right) B_\nu$, one simply gets
\be
TD_3=0\quad.\label{en7}
\ee

\item
From $\omega^{\mu\nu}B_\nu$, one has
\be
\left(T\CD_\mu\bar\mu -{1\over n}\CD_\mu p\right)\left(-\xi_3 +{\sigma\over n}\left(\xi-2TD_B\right)\right)
+T\CD_\mu D_3=0\quad,
\ee
and using the fact that $D_3=0$ from above, one simply gets
\be
\xi_3={\sigma\over n}\left(\xi-2 T D_B\right)\quad,
\ee
or equivalently
\be
\bar\xi_3 = {\bar\sigma \over \bar n}\left(\bar\xi-2\bar D_B\right)\quad,\label{en8}
\ee
writing $\xi_3=\bar \xi_3$, $\sigma=T\bar\sigma$ according to the conformal weights.

\item From $\sigma_{\mu\nu}\CD^\mu\omega^\nu$, one arrives at
\be
-\lambda_1+T D_1=0\quad,
\ee
or upon writing $\lambda_1=T^2\bar\lambda_1$,
\be
\bar\lambda_1=\bar D_1\quad.\label{en9}
\ee

\item From $\sigma_{\mu\nu}\CD^\mu B^\nu$, it leads to
\be
-\lambda_4+TD_2 =0\quad,\quad {\rm equivalently}\quad \bar\lambda_4 = \bar D_2\quad,\label{en10}
\ee
where $\lambda_4=T\bar\lambda_4$.

\item $\omega_{\mu\nu}\CD^\mu B^\nu$ gives one the same equation as in (\ref{en7}),
\be
TD_3=0\quad.
\ee

\item 
From $\epsilon^{\mu\nu\alpha\beta}\omega_{\mu\nu}\CD_\alpha E_\beta$, one simply concludes that 
\be
TD_5 =0\quad.\label{en11}
\ee

\item What remains can be grouped into two components. One is proportional to \linebreak $\epsilon^{\mu\nu\alpha\beta}u_\nu\CD_\alpha p\CD_\beta\bar\mu$ and the other is proportional to
$\epsilon^{\mu\nu\alpha\beta}u_\nu\CD_\alpha\left({1\over n}\CD_\beta p\right)$. 
Now, the latter can be expanded as
\be
\epsilon^{\mu\nu\alpha\beta}u_\nu\CD_\alpha\left({1\over n}\CD_\beta p\right)=-{1\over n^2}\epsilon^{\mu\nu\alpha\beta}u_\nu \CD_\alpha n\CD_\beta p+{2p\over n}\epsilon^{\mu\nu\alpha\beta}u_\nu\CW_{\alpha\beta}\quad,
\ee
and the second piece is completely independent of $\epsilon^{\mu\nu\alpha\beta}u_\nu\CD_\alpha p\CD_\beta\bar\mu$, so that one can treat the above two components $\epsilon^{\mu\nu\alpha\beta}u_\nu\CD_\alpha p\CD_\beta\bar\mu$ and  $\epsilon^{\mu\nu\alpha\beta}u_\nu\CD_\alpha\left({1\over n}\CD_\beta p\right)$ independently;
each component should then vanish separately. The first component gives
\be
{T\over n}\CD_\mu D_4 =0\quad,
\ee
and since $D_4=\bar D_4$, it leads to 
\be
\bar D_4'=0\quad.\label{en12}
\ee
The coefficient of the second component leads to
\be
-\xi_5\left(T\CD_\mu\bar\mu-{1\over n}\CD_\mu p\right)+T\CD_\mu D_5+TD_4\CD_\mu\bar\mu
+{T\over 4p}D_5 \CD_\mu p=0\quad,
\ee
and using (\ref{en11}) $D_5=0$, this leads to two equations
\bear
\bar D_4=0\quad,\quad \bar\xi_5 =0\quad,\label{en13}
\eear
where $\xi_5=\bar\xi_5$.
This completes the list of all contraints derived from the requirement of no entropy production.

\end{itemize}

Let us now solve these equations in a more explicit form. Recall that the first order coefficients $\bar\xi$, $\bar\xi_B$, $\bar D_B$ are already given in the previous subsection in (\ref{firres1}) and (\ref{firres2}).
Equations (\ref{en3}), (\ref{en4}), (\ref{en7}), (\ref{en8}), (\ref{en9}), (\ref{en10}), (\ref{en11}), and (\ref{en13}) trivially give solutions for $\bar D_1$, $\bar D_2$, $\bar D_3$, $\bar\xi_3$, $\bar\lambda_1$, $\bar\lambda_4$, $\bar D_5$, $\bar D_4$, and $\bar\xi_5$ as
\bear
&&\bar D_1= \bar\lambda_1= {2\bar\eta\over\bar n}\left(\bar\xi-2\bar D_B\right)\,\,,\,\,
\bar D_2= \bar\lambda_4= {2\bar\eta\over\bar n}\left(\bar\xi_B-\kappa\bar\mu\right)\,\,,\,\, \bar\xi_3={\bar\sigma\over \bar n} \left(\bar\xi-2\bar D_B\right)\,\,,
\nonumber\\
&&\bar D_3=\bar D_4=\bar D_5=\bar\xi_5=0\quad.
\eear
From (\ref{en1}), (\ref{en2}) and using the solution for $\bar D_1$, one can solve
\bear
\bar\lambda_2+\bar\xi_1&=&\left({2\bar\eta\over\bar n}\left(\bar\xi-2\bar D_B\right)
\right)'+\left({\bar\eta\over\bar p}-{2\bar\eta'\over\bar n}\right)\left(\bar\xi-2\bar D_B\right)\quad,\\
\bar\lambda_8-\bar\xi_1&=&-{\bar\eta\over \bar p}\left(\bar\xi-2\bar D_B\right)\quad,
\eear
which fixes $\bar\lambda_2+\bar\xi_1$ and $\bar\lambda_8-\bar\xi_1$ leaving
one unknown such as $\bar\lambda_2-\bar\xi_1$.
Finally, the equations (\ref{en5}) and (\ref{en6}) with the solution for $\bar D_2$ give
\bear
\bar\lambda_5+\bar\xi_2&=& \left({2\bar\eta\over\bar n}\left(\bar\xi_B-\kappa\bar\mu\right)\right)'
+\left({3\over 2}{\bar\eta\over \bar p}-{2\bar\eta'\over\bar n}\right)\left(\bar\xi_B-\kappa\bar\mu\right)\quad,\\
\bar\lambda_6-\bar\xi_2&=& -{3\over 2}{\bar\eta\over\bar p}\left(\bar\xi_B-\kappa\bar\mu\right)\quad,
\eear
leaving $\bar\lambda_5-\bar\xi_2$ undertermined. In summary, we have determined thirteen 
out of eighteen anomalous second order transport coefficients, leaving five parameters $\bar\lambda_3$, $\bar\lambda_2-\bar\xi_1$, $\bar\lambda_5-\bar\xi_2$, $\bar\lambda_7$, and $\bar\xi_4$ unfixed by our method.

Refs.\cite{Erdmenger:2008rm,Banerjee:2008th} computed four transport coefficients of  interest for us, $(\lambda_{1,2,3},\xi_1)$ in the framework of AdS/CFT correspondence, explicitly corresponding to the $N=4$ super Yang-Mills theory with $U(1)$ R-symmetry at strong coupling. 
One can test some of our relations against these computations; specifically, one can test two relations that we have derived,
\bear
\bar\lambda_1&=& {2\bar\eta\over\bar n}\left(\bar\xi-2\bar D_B\right)\quad,\nonumber\\
\bar\lambda_2+\bar\xi_1&=&\left({2\bar\eta\over\bar n}\left(\bar\xi-2\bar D_B\right)
\right)'+\left({\bar\eta\over\bar p}-{2\bar\eta'\over\bar n}\right)\left(\bar\xi-2\bar D_B\right)\quad,\label{test}
\eear
which are quite non-trivial.
Let us first summarize the results of Refs.\cite{Erdmenger:2008rm,Banerjee:2008th}, especially following notations in Ref.\cite{Erdmenger:2008rm}.
The 5D action is
\be
(16\pi G_5){\cal L}_{5D} = R+12-{1\over 4}F_{MN}F^{MN} - {1\over 12\sqrt{3}}{\epsilon^{MNPQR}\over\sqrt{-g}}
A_M F_{NP} F_{QR}\quad,\label{5d}
\ee
where $G_5$ is related to the gauge theory by $G_5={\pi\over 2N_c^2}$.
The charged black hole solution is
\bear
ds^2&=& -r^2f(r) dt^2 +2 dt dr + r^2\sum_{i=1,2,3} \left(dx^i\right)^2\quad,\quad f(r)=1-{m\over r^4}+{Q^2\over 3r^6}\quad,\nonumber\\
A&=& -{Q\over r^2} dt\quad,\label{bhsol}
\eear
The temperature $T$ and the chemical potential $\mu$ are given in terms of parameters $(m,Q)$ by
\be
T={r_H^2f'(r_H)\over 4\pi}\quad,\quad \mu={Q\over r_H^2}\quad,
\ee
where the horizon $r_H$ is the largest solution of $f(r_H)=0$.
These can be solved explicitly as \cite{Erdmenger:2008rm}
\be
\bar r_H\equiv {r_H\over T}={\pi\over 2}\left(1+\sqrt{1+{2\bar\mu^2\over 3\pi^2}}\right)\,,\,
\bar m\equiv {m\over T^4}={\pi^4\over 16}\left(1+\sqrt{1+{2\bar\mu^2\over 3\pi^2}}\right)^3\left(3\sqrt{1+{2\bar\mu^2\over 3\pi^2}}-1\right)\,,
\ee
in terms of which the relevant quantities are given by
\bear
&&\bar p = {\bar m\over 16\pi G_5}\quad,\quad \bar n = {\bar r_H^2\bar\mu\over 8\pi G_5}\quad,\quad
\bar\eta= {\bar r_H^3\over 16\pi G_5}\quad,\nonumber\\
&&\bar\xi={\bar r_H^4\bar\mu^2\over 8\sqrt{3}\pi G_5 \bar m}\quad,\quad
\bar D_B={\bar\mu^2\over 16\sqrt{3}\pi G_5}\quad,\quad
\bar\lambda_1=-{\bar r_H^3\bar\mu^3\over 24\sqrt{3}\pi G_5 \bar m}\quad,\nonumber\\
&&\bar\lambda_2=\bar\lambda_3 =0\quad,\quad\bar\xi_1=-{\bar r_H^7\bar\mu^2\over 8\sqrt{3}\pi G_5 \bar m^2}\quad.
\eear
It is satisfying to see that the two relations (\ref{test}) as well as $\bar p'=\bar n$
are obeyed by the above holographic results; this is a rather non-trivial test of our guiding principle. 
It would be interesting to perform a full-fledged fluid/gravity correspondence computation
for other transport coefficients we identify and to check other relations too.

\subsection{Chiral shear wave} 

In this subsection, we discuss one physics phenomenon that is related to the second order
viscous corrections from triangle anomaly addressed in the previous subsections: the chiral shear wave.
The chiral shear wave is a modification of the transverse shear mode dispersion relation 
\be
\omega \approx -i{\eta\over  \epsilon+p} k^2 \pm i C k^3+  \cdots\quad,
\ee
where the leading $k^2$ piece is as usual, and the $k^3$ term is the first effect originated from anomaly.
If the system is parity invariant without triangle anomaly, the dispersion relation should be
invariant under parity transformation $k\to-k$, so that only even powers of $k$ should have appeared
in the dispersion relation. Any odd powers of $k$ are effects from triangle anomaly, and the $k^3$ term
in the above is the first of them. The $\pm$ sign in front of it depends on the helicity of the shear modes which will become clear shortly. 

The chiral shear wave was first observed in Refs.\cite{Sahoo:2009yq,Matsuo:2009xn,Nakamura:2009tf} via AdS/CFT correspondence of $N=4$ super Yang-Mills with $U(1)$ R-symmetry at strong coupling. The 5D action and the charged black hole solution which serves as a background are precisely same as in the last subsection, (\ref{5d}) and (\ref{bhsol}).
The resulting shear mode dispersion relation was \cite{Sahoo:2009yq}
\be
\omega\approx -i{\eta\over 4p}k^2 \pm i{r_H^3\mu^3 \over 24\sqrt{3} m^2} k^3 +\cdots\quad.\label{cxw}
\ee
We would like to understand the origin of the $k^3$ piece in terms of our second order viscous corrections
from triangle anomaly to the energy-momentum tensor and the $U(1)$ current. One can easily check that
the first order corrections do not induce chiral shear wave, and it has been expected that it should be related to higher order corrections \footnote{At least to the authors, it was first pointed out by D.T.Son. Ref.\cite{Amado:2011zx} also mentioned it recently.}.
We will see that it actually comes from second order corrections we identify.

To derive dispersion relations of linearized fluctuations, one starts with a static background fluid
of temperature $T$ and chemical potential $\mu$, and considers small fluctuations of hydrodynamic variables; in our case $\delta T$, $\delta\bar\mu$, and $\delta u^\mu$, and keep only terms that are linear in their amplitudes when one writes $\delta T^{\mu\nu}$ and $\delta j^\mu$.
For our purpose we don't have external electric/magnetic fields, so that the hydrodynamic equations of motion
\be
\partial_\mu \delta T^{\mu\nu}=0\quad,\quad \partial_\mu\delta j^\mu =0\quad,
\ee
will give us spacetime propagation of these linearized fluctuations. In the frequency-momentum space, we can read off dispersion relations.

One first picks up a definite frequency-momentum $(\omega,\vec k = k \hat x^1)$, or equivalently every fluctuating mode is assumed to have a common phase factor $e^{-i\omega t +i k x^1}$. 
Due to a residual $SO(2)$ symmetry in the transverse $(x^2,x^3)$ space, fluctuating modes are classified
by their helicities under $SO(2)$ rotation, and different helicity modes do not mix.
We are interested in helicity $\pm 1$ transverse shear fluctuations, 
\be
\delta u_{\pm 1}=\left(\delta u^2\pm i \delta u^3\right)\quad.
\ee
Note that they are the only modes with this helicity because other possible modes such as
$\partial_{2,3}( \delta T,\delta\bar\mu)=0$ are simply absent because $\partial_{2,3}=0$ by our assumption of momentum direction. Therefore, one only needs to consider fluctuations $\delta u_{\pm 1}$.
Because we know that $k^3$ term in the dispersion relation is coming from anomaly and the usual $k^2$ term from the first order correction, we can neglect non-anomalous second order corrections to the energy-momentum tensor and the $U(1)$ current for our purpose. Out of the eight anomalous second order corrections to the energy-momentum tensor, one easily sees that only the term $
\lambda_1 \Pi^{\mu\nu}_{\alpha\beta}\CD^\alpha\omega^\beta$ gives non-zero contributions at our linearized level, so that one needs to compute
\be
\delta T^{\mu\nu} = 4p \delta u^\mu u^\nu+4p u^\mu\delta u^\nu -2\eta\delta\sigma^{\mu\nu}
+\lambda_1 \delta\left(\Pi^{\mu\nu}_{\alpha\beta}\CD^\alpha\omega^\beta\right)\quad,
\ee
whose non-vanishing components after some algebra are
\bear
\delta T^{02}&=& 4p\delta u^2 +{\lambda_1\over 4}k\omega \delta u^3\quad,\nonumber\\
\delta T^{03}&=& 4p\delta u^3 -{\lambda_1\over 4}k\omega \delta u^2\quad,\nonumber\\
\delta T^{12}&=& -i\eta k \delta u^2 +{\lambda_1\over 4}k^2 \delta u^3\quad,\nonumber\\
\delta T^{13}&=& -i\eta k \delta u^3 -{\lambda_1\over 4}k^2 \delta u^2\quad.
\eear
Then, the energy-momentum conservation gives
\bear
\partial_0 T^{02}+\partial_1 T^{12}&=&-i\omega\left(4p\delta u^2 +{\lambda_1\over 4}k\omega \delta u^3\right)
+ik\left(-i\eta k \delta u^2 +{\lambda_1\over 4}k^2 \delta u^3\right)=0\quad,\nonumber\\
\partial_0 T^{03}+\partial_1 T^{13}&=&-i\omega\left(4p\delta u^3 -{\lambda_1\over 4}k\omega \delta u^2\right)
+ik\left(-i\eta k \delta u^3 -{\lambda_1\over 4}k^2 \delta u^2\right)=0\quad,\nonumber\\
\eear
and in terms of $\delta u_{\pm 1}= \left(\delta u^2\pm i \delta u^3\right)$, they become
\bear
\left(-i\omega\left(4p\mp i {\lambda_1\over 4}k\omega\right)
+ik\left(-i\eta k  \mp i {\lambda_1\over 4} k^2\right)\right)\delta u_{\pm 1}=0\quad,\\
\eear
which is solved up to order $k^3$ as
\be
\omega \approx -i{\eta\over 4p}k^2 \mp i {\lambda_1\over 16p} k^3 +\cdots\quad.
\ee
We see that the $k^3$ term indeed originates from one of our second order transport coefficients $\lambda_1$.

Let us confirm this in the AdS/CFT computation: $\lambda_1$ and $p$ as given by the AdS/CFT computations are given in the previous subsection, and recalling $\lambda_1=T^2\bar\lambda_1$, $p=T^4\bar p$, one can check that
\be
-{\lambda_1\over 16p}=-{\bar\lambda_1\over 16 T^2\bar p}=
{\bar r_H^3\bar\mu^3 \over 24\sqrt{3} T^2\bar m^2 } = {r_H^3\mu^3\over 24\sqrt{3} m^2}\quad,
\ee
which agrees with the independent result (\ref{cxw}) from the linearized analysis in Ref.\cite{Sahoo:2009yq}.

\section{Hydrodynamics with anomaly in higher dimensions}\label{part2}

\subsection{Entropy current method}

In this second part of the paper, we will relax the constraint of conformal symmetry, but instead of staying in four
spacetime dimensions, we will explore higher even $2N$ dimensional spacetime with $(N+1)$-gon anomaly
among multiple $U(1)$ symmetries. We will follow the entropy production constraint similar to the one used in Ref.\cite{Son:2009tf}
and in the first part of this work. The effects from the anomaly first appear at $(N-1)$'th order in derivative expansion, and
we will identify $N$ number of terms in the $U(1)$ currents that originate
from anomaly. We will see that requiring positivity of entropy production fixes all the transport coefficients corresponding to them. 

For a concise presentation, we will show details of our derivation for the case of a single $U(1)$ symmetry;
the results for the multiple $U(1)$ case follow from a straightforward generalization of it.
The starting point is again the basic hydrodynamic equations
\bear
\nabla_\mu T^{\mu\nu}&=& F^{\nu\alpha}j_\alpha\quad,\nonumber\\
\nabla_\mu j^\mu &=& {\kappa\over 2N^2}\epsilon^{\mu_1\nu_1\cdots\mu_{N}\nu_{N}}F_{\mu_1\nu_1}\cdots F_{\mu_{N}\nu_{N}}= \kappa E_\mu\, B^\mu_{(0,N-1)}\quad,\label{basic2}
\eear
where we define
\bear
E^\mu&=& F^{\mu\nu} u_\nu\quad,\nonumber\\
B^\mu_{(0,N-1)}&=& {1\over N}\epsilon^{\mu\nu\alpha_1\beta_1\cdots\alpha_{N-1}\beta_{N-1}}u_\nu F_{\alpha_1\beta_1}\cdots F_{\alpha_{N-1}\beta_{N-1}}\quad.
\eear
The meaning of the subscript $(0,N-1)$ in $B^\mu_{(0,N-1)}$ will become clear in a moment.
One then invokes derivative expansion in writing down constitutive relations for $T^{\mu\nu}$ and $j^\mu$
in terms of thermodynamic variables of plasma in the Landau frame as
\bear
 T^{\mu\nu}& =& \left(\epsilon+p\right)u^\mu u^\nu +p g^{\mu\nu} +\tau^{\mu\nu}\quad,\nonumber\\
j^\mu&=& n u^\mu +\nu^\mu\quad,
\eear
where the viscous corrections $\tau^{\mu\nu}$ and $\nu^\mu$ are transverse by the definition of Landau frame,
\be
u_\mu\tau^{\mu\nu}= u_\mu \nu^\mu =0\quad.
\ee
We assume that the basic thermodynamic relations hold:
\bear
\epsilon+p=Ts+\mu n\quad,\quad d\epsilon = T ds+\mu dn\quad,\quad dp=s dT+n d\mu\quad,
\eear
where $s$ is the entropy density and $\mu$ is the chemical potential.
As is by now a standard procedure, one considers
\bear
&&u_\nu \nabla_\mu T^{\mu\nu}+\mu\nabla_\mu j^\mu = -E_\mu \nu^\mu+\kappa\mu E_\mu B^\mu_{(0,N-1)}\nonumber\\
&&= -u^\mu\nabla_\mu \epsilon+\mu u^\mu\nabla_\mu n-\left(\epsilon+p -\mu n\right)\nabla_\mu u^\mu
+u_\nu\nabla_\mu\tau^{\mu\nu}+\mu \nabla_\mu \nu^\mu\nonumber\\
&&=-T u^\mu\nabla_\mu s -Ts \nabla_\mu u^\mu +u_\nu\nabla_\mu\tau^{\mu\nu}+\mu \nabla_\mu \nu^\mu\nonumber\\
&&=-T\nabla_\mu\left(su^\mu -\bar\mu \nu^\mu \right)-\left(\nabla_\mu u_\nu\right) \tau^{\mu\nu}
-T\left(\nabla_\mu\bar\mu\right)\nu^\mu\quad,
\eear
where $\bar\mu\equiv {\mu\over T}$. This leads one to
\bear
T\nabla_\mu\left(s u^\mu-\bar\mu\nu^\mu\right) = -\left(\nabla_\mu u_\nu\right)\tau^{\mu\nu}
-\left( T \nabla_\mu\bar\mu-E_\mu \right) \nu^\mu -\kappa\mu E_\mu B^\mu_{(0,N-1)}\quad,\label{st2}
\eear
which is a typical starting point for the consideration of entropy production.

We consider $F_{\mu\nu}$ as being first order in derivative, therefore the anomaly term in the basic hydrodynamic equation (\ref{basic2}) is of $N$'th order in derivative. The left-hand side is $\nabla_\mu j^\mu$, so it is expected that the first effects from anomaly should appear in $j^\mu$ at $(N-1)$'th
order in derivative. This is a generalization of the case of $N=2$ or in four dimensions where indeed the triangle anomaly affects $j^\mu$ at $(N-1)=1$st order. The corrections to the energy-momentum tensor
are expected to be of higher order, presumably starting at $N$'th order in derivative, which is beyond of our interest in this section.
We stress that there should in general be many other non-anomalous viscous terms of
$(N-1)$'th or less order in derivative, and we cannot possibly classify them all.
However, the point is that the terms that are necessarily linear in the anomaly coefficients $\kappa$
do not mix with these other non-anomalous terms due to discrete symmetries $(C,P)$ that we discuss in the previous section, so that it makes sense to discuss them separately. In $2N$ dimensions, the $(C,P)$ charges 
are similar to those in four dimensions in the previous section, except $B^\mu_{(0,N-1)}$ now has
$(C,P)=((-1)^{N-1},+1)$ and $\kappa$ has $(C,P)=((-1)^{N-1},-1)$, so that the transport coefficients are classified by there $(C,P)$ charges as follows,
\bear
(C,P)=(+1,+1)\quad&:&\quad f(T,\bar\mu,\kappa)= g(T,\bar\mu^2,\kappa^2)\nonumber\\
(C,P)=(-1,+1)\quad&:&\quad f(T,\bar\mu,\kappa)=\bar\mu g(T,\bar\mu^2,\kappa^2)\nonumber\\
(C,P)=((-1)^{N-1},-1)\quad&:&\quad f(T,\bar\mu,\kappa)= \kappa g(T,\bar\mu^2,\kappa^2)\nonumber\\
(C,P)=((-1)^N,-1)\quad&:&\quad f(T,\bar\mu,\kappa)= \kappa\bar\mu g(T,\bar\mu^2,\kappa^2)\nonumber
\eear
To summarize, the transport coefficients of $P=-1$ that are of interest for us are necessarily linear in $\kappa$.

When one constructs the viscous corrections from various derivatives of thermodynamic quantities, the basic
hydrodynamic equations (\ref{basic2}) can be used to remove some of the first order derivative terms up to higher order terms, so that
one does not need them in constructing viscous terms at a given fixed order in derivative.
As the total number of equations in (\ref{basic2}) is $(N+1)$, one expects to be able to remove
$(N+1)$ first order derivative terms using the equations of motion. The general
first order derivative terms are $\nabla_\mu u^\nu$, $F_{\mu\nu}$ and derivatives of any two independent thermodynamic scalars, say $\nabla_\mu(T,\bar\mu)$ or $\nabla_\mu(p,n)$, because locally the plasma is completely specified by two independent
thermodynamic scalars. It is a matter of choice which $(N+1)$ terms among the above first derivative terms
are removed by using the equations of motion (\ref{basic2}). For our convenience, we will remove
\be
\nabla_\mu p\quad,\quad u^\mu\nabla_\mu n\quad,
\ee
so that the remaining available building blocks of constructing viscous terms are simply
\be
\nabla_\mu u^\nu\quad,\quad F_{\mu\nu}\quad,\quad\Delta^{\mu\nu}\nabla_\nu n\quad,
\ee
where $\Delta^{\mu\nu}=u^\mu u^\nu+g^{\mu\nu}$ is the projection operator
to the space transverse to $u^\mu$.

To make things explicit, let us work out in detail how the above mentioned removal happens.
Up to higher order derivative terms, the basic equations of motion (\ref{basic2}) are written as
\bear
\nabla_\mu T^{\mu\nu}&=&u^\mu\nabla_\mu \left(\epsilon+p\right)u^\nu +\left(\epsilon+p\right)\left(\nabla_\mu u^\mu\right)u^\nu +\left(\epsilon+p\right)u^\mu\nabla_\mu u^\nu+\nabla^\nu p = n E^\nu\quad,\nonumber\\
\nabla_\mu j^\mu &=& u^\mu\nabla_\mu n+n\nabla_\mu u^\mu =0\quad.\label{b2}
\eear
From the second equation, one has
\be
u^\mu\nabla_\mu n = -n \left(\nabla_\mu u^\mu\right)\quad.\label{rem1}
\ee
On the other hand, contracting the first equation with $u_\nu$, one gets
\be
\left(\epsilon+p\right)\left(\nabla_\mu u^\mu\right)=-u^\mu\nabla_\mu \epsilon \quad,\label{b3}
\ee
and inserting this into the first equation of (\ref{b2}), one can obtain after an easy  manipulation
\be
\Delta^{\nu\mu}\nabla_\mu p = n E^\nu -\left(\epsilon+p\right) u^\mu\nabla_\mu u^\nu\quad.\label{rem2}
\ee
Finally, from (\ref{b3}) and writing $\epsilon=\epsilon(p,n)$ and using (\ref{rem1}), 
\bear
-\left(\epsilon+p\right) \left(\nabla_\mu u^\mu\right) &=&u^\mu\nabla_\mu\epsilon=\left(d\epsilon\over dp\right)_n u^\mu \nabla_\mu p+
\left(d\epsilon\over dn\right)_p u^\mu \nabla_\mu n\nonumber\\
&=& \left(d\epsilon\over dp\right)_n u^\mu \nabla_\mu p-
n\left(d\epsilon\over dn\right)_p \left(\nabla_\mu u^\mu\right)\quad,
\eear
so that one finally arrives at
\be
u^\mu\nabla_\mu p = {\left(n\left(d\epsilon\over dn\right)_p -\left(\epsilon+p\right)\right)\over 
\left(d\epsilon\over dp\right)_n }\left(\nabla_\mu u^\mu\right)\quad.\label{rem3}
\ee
The expressions (\ref{rem1}), (\ref{rem2}), and (\ref{rem3}) indeed replace
$\nabla_\mu p$ and $u^\mu\nabla_\mu n$ with $\nabla_\mu u^\nu$ and $F_{\mu\nu}$ up to higher order derivatives, so that we can remove them.

We are interested in the anomaly-induced viscous corrections to the current $\nu^\mu$. Since $\nu^\mu$ is a vector
whose spatial component has $P=-1$ and we want a transport coefficient having $P=-1$ to be linear in $\kappa$,
any anomalous viscous correction to $\nu^\mu$ should be a pseudo-vector whose spatial component has $P=+1$.
It is easy to see that the first possible pseudo-vectors one can construct out of $\nabla_\mu u^\nu$, $F_{\mu\nu}$, $\Delta^{\mu\nu}\nabla_\nu n$ and their derivatives indeed start to appear
at $(N-1)$'th order in derivative, containing one $\epsilon$-tensor. It is also not difficult to see that the 
anomalous corrections to the energy-momentum tensor appear only at $N$'th order and beyond.
At $(N-1)$'th order of our interest, we find precisely $N$ possible pseudo-vectors which
can appear in $\nu^\mu_{(N-1)}$ and $s^\mu_{(N-1)}$;
\be
B^\mu_{(s,t)}={1\over N}\epsilon^{\mu\nu\alpha_1\beta_1\cdots\alpha_s\beta_s \gamma_1\delta_1\cdots\gamma_t\delta_t}u_\nu \left(\nabla_{\alpha_1} u_{\beta_1}\right)
\cdots\left(\nabla_{\alpha_s} u_{\beta_s}\right) F_{\gamma_1\delta_1}
\cdots F_{\gamma_t\delta_t}\quad,\quad s+t=N-1\quad.
\ee
The $B^\mu_{(0,N-1)}$ and $B^\mu_{(N-1,0)}$ are $2N$-dimensional generalizations of
$B^\mu ={1\over 2}\epsilon^{\mu\nu\alpha\beta}u_\nu F_{\alpha\beta}$ and $\omega^\mu=
{1\over 2}\epsilon^{\mu\nu\alpha\beta}u_\nu \left(\nabla_\alpha u_\beta\right)$ in four dimensions, while other $B^\mu_{(s,t)}$ with $0<s<(N-1)$ exist only in higher dimensions.
One then introduces $2N$ transport coefficients $\xi_{(s,t)}$ and $D_{(s,t)}$ as
\bear
\nu^\mu_{(N-1),{\rm anomaly}}=\sum_{s+t=N-1}\xi_{(s,t)} B^\mu_{(s,t)}\quad,\quad
s^\mu_{(N-1),{\rm anomaly}}=\sum_{s+t=N-1}D_{(s,t)} B^\mu_{(s,t)}\quad,\label{transport2}
\eear
and inserts these into the entropy production formula (\ref{st2}) to get some constraints on them.

To proceed, we need the following formula derived from using the equations of motion:
\bear
\nabla_\mu B^\mu_{(s,t)} = -{(s+1)\over\left(\epsilon+p\right)} B^\mu_{(s,t)} \left(\nabla_\mu p - n E_\mu\right)
-2(N-1-s) E_\mu B^\mu_{(s+1,t-1)}\quad,\label{midrel}
\eear
where by definition $B^\mu_{(s,t)}=0$ if $s\ge N$, and the above equality holds true up to higher order corrections.

{\it Proof : } Let us start from the definition
\bear
\nabla_\mu B^\mu_{(s,t)} ={1\over N}\epsilon^{\mu\nu\alpha_1\beta_1\cdots\gamma_1\delta_1\cdots}\left(\nabla_\mu u_\nu\right)\left(\nabla_{\alpha_1} u_{\beta_1}\right)\cdots F_{\gamma_1\delta_1}\cdots\quad,
\eear
where other possible actions of $\nabla_\mu$ to $\left(\nabla_{\alpha_i} u_{\beta_i}\right)$ or $ F_{\gamma_j\delta_j}$ give simply zero using Bianchi identities of the Riemann tensor and the field strength tensor. It is most convenient to work in the local rest frame where $u^i=0$ ($i=1,2,3$) and $\nabla_\mu u_0=0$,
so that the above becomes
\bear
\nabla_\mu B^\mu_{(s,t)} &=&{(s+1)\over N}\epsilon^{0\nu\alpha_1\beta_1\cdots\gamma_1\delta_1\cdots}\left(\nabla_0 u_\nu\right)\left(\nabla_{\alpha_1} u_{\beta_1}\right)\cdots F_{\gamma_1\delta_1}\cdots\nonumber\\
&+&{2(N-1-s)\over N}\epsilon^{\mu\nu\alpha_1\beta_1\cdots0\delta_1\cdots}\left(\nabla_\mu u_\nu\right)\left(\nabla_{\alpha_1} u_{\beta_1}\right)\cdots F_{0\delta_1}\cdots\quad.\label{mid1}
\eear
Next, multiplying the equation of motion $\nabla_\mu T^{\mu\nu}$ by $B_{\nu(s,t)}$ 
and using that $u^\nu B_{\nu(s,t)}=0$, one obtains  up to higher derivitive corrections the following relation
\bear
{1\over N}\left(\epsilon+p\right)u^\mu u_\lambda
\epsilon^{\nu\lambda\alpha_1\beta_1\cdots\gamma_1\delta_1\cdots}\left(\nabla_\mu u_\nu\right)
\left(\nabla_{\alpha_1} u_{\beta_1}\right)\cdots F_{\gamma_1\delta_1}\cdots=B^\mu_{(s,t)}\left(-\nabla_\mu p+n E_\mu\right),
\eear
which becomes in the local rest frame
\bear
{1\over N}\left(\epsilon+p\right)
\epsilon^{0\nu\alpha_1\beta_1\cdots\gamma_1\delta_1\cdots}\left(\nabla_0 u_\nu\right)
\left(\nabla_{\alpha_1} u_{\beta_1}\right)\cdots F_{\gamma_1\delta_1}\cdots=B^\mu_{(s,t)}\left(-\nabla_\mu p+n E_\mu\right).\label{mid2}
\eear
The left-hand side is precisely proportional to the first term in (\ref{mid1}).
Finally, using $F_{0\delta_1}=u_0 E_{\delta_1}$ one manipulates the second term in (\ref{mid1}) as
\bear
&&\epsilon^{\mu\nu\alpha_1\beta_1\cdots0\delta_1\cdots}\left(\nabla_\mu u_\nu\right)\left(\nabla_{\alpha_1} u_{\beta_1}\right)\cdots F_{0\delta_1}\cdots\nonumber\\
&&=u_0E_{\delta_1}\epsilon^{0\delta_1\mu\nu\alpha_1\beta_1\cdots \gamma_2\delta_2\cdots}
\left(\nabla_\mu u_\nu\right)\left(\nabla_{\alpha_1} u_{\beta_1}\right)\cdots F_{\gamma_2\delta_2}\cdots\label{mid3}\\
&&=-E_{\delta_1}\epsilon^{\delta_1 0\mu\nu\alpha_1\beta_1\cdots \gamma_2\delta_2\cdots}
u_0\left(\nabla_\mu u_\nu\right)\left(\nabla_{\alpha_1} u_{\beta_1}\right)\cdots F_{\gamma_2\delta_2}\cdots=
-NE_{\delta_1}B^{\delta_1}_{(s+1,t-1)}\quad.\nonumber
\eear
The (\ref{mid1}), (\ref{mid2}), and (\ref{mid3}) prove our relation (\ref{midrel}). (QED)

Using (\ref{midrel}), the entropy production formula (\ref{st2}) at $(N-1)$'th order upon inserting (\ref{transport2})
becomes
\bear
&&-\left( T \nabla_\mu\bar\mu-E_\mu \right) \nu^\mu_{(N-1),{\rm anomaly}} -\kappa\mu E_\mu B^\mu_{(0,N-1)}
+T\nabla_\mu s^\mu_{(N-1),{\rm anomaly}}\nonumber\\
&&=-T \sum_{s+t=N-1}\left(\left(\nabla_\mu\bar\mu\right)\xi_{(s,t)}-\nabla_\mu D_{(s,t)}
+{(s+1)\over\left(\epsilon+p\right)}\left(\nabla_\mu p\right) D_{(s,t)}\right) B^\mu_{(s,t)}\nonumber\\
&&+\sum_{s+t=N-1}
\left(\xi_{(s,t)}+{(s+1)Tn\over\left(\epsilon+p\right)}  D_{(s,t)}-2T(N-s)D_{(s-1,t+1)}\right)E_\mu B^\mu_{(s,t)}\quad,
\eear
where we formally define
\be
D_{(-1,N)} = {1\over 2N}\kappa\bar\mu\quad.
\ee

It is important to remember that the equations of motion relate $E^\nu$ to
\be
E^\nu={1\over n}
\left(\Delta^{\nu\mu}\nabla_\mu p +\left(\epsilon+p\right) u^\mu\nabla_\mu u^\nu\right) \quad, 
\ee
using (\ref{rem2}), so one needs to be a bit careful when considering arbitrary possible independent configurations. The above entropy formula has a structure
\be
\left(A_{\mu(s,t)}+C_{(s,t)}E_\mu\right) B^\mu_{(s,t)}\quad,
\ee
and the coefficients $A_{\mu(s,t)}$, $C_{(s,t)}$ involve only thermodynamic scalars and their derivatives. As $B^\mu_{(s,t)}$ are clearly arbitrary and independent, the total coefficients in front of each of them should vanish 
in order to make sure positivity of entropy production, so the basic constraints one derives is in fact
\be
A_{\mu(s,t)}+C_{(s,t)}E_\mu=0\quad.\label{cot}
\ee
An important observation is that $E_\mu$ contains $u^\nu\nabla_\nu u_\mu$-piece while 
$A_{\mu(s,t)}$, $C_{(s,t)}$ are made of thermodynamic scalars and their derivatives only, without any $u^\nu\nabla_\nu u_\mu$. Because $u^\nu\nabla_\nu u_\mu$ can vary independently of derivatives of thermodynamic scalars,
imposing (\ref{cot}) for arbitrary possible configurations gives us a simpler conclusion that
\be
A_{\mu(s,t)}=0\quad,\quad C_{(s,t)}=0\quad,\label{cot2}
\ee
separately. 

The reason why we expound on this subtlety is due to an interesting difference from the conformal case that we have studied in the previous section.
Recall that in that case, $E_\mu ={1\over n}\CD_\mu p$ by equations of motion and the $A_\mu$ and $C$
are also expressed in terms of $\CD_\mu$ of thermodynamic variables, so that one can no longer consider
$E_\mu$ as being independent, and the constraint one has is only (\ref{cot}), not (\ref{cot2}) anymore.
What happens in the conformal case is that $u^\nu\nabla_\nu u_\mu$ and $\nabla_\mu (T,\bar\mu)$
are packaged to give only two independent combinations $\CD_\mu(T,\bar\mu)$, not the general three,
so that the number of constraints one gets is reduced. However, what saves us is the conformal symmetry constraint on the transport coefficients $f=T^w\bar f(\bar\mu)$ which is given from the start,
so that one is still able to solve the system of reduced constraints, as we do in the previous section.

Therefore, one has the system of $2N$ algebro-differential equations
\bear
\left(\nabla_\mu\bar\mu\right)\xi_{(s,t)}-\nabla_\mu D_{(s,t)}
+{(s+1)\over\left(\epsilon+p\right)}\left(\nabla_\mu p\right) D_{(s,t)}&=&0\quad,\label{diff1}\\
\xi_{(s,t)}+{(s+1)Tn\over\left(\epsilon+p\right)}  D_{(s,t)}-2T(N-s)D_{(s-1,t+1)}&=&0\quad,\label{diff2}
\eear
with the boundary condition $D_{(-1,N)} = {1\over 2N}\kappa\bar\mu$. We can solve them  analytically.
Guided by Ref.\cite{Son:2009tf}, one chooses $(p,\bar\mu)$ as the two independent thermodynamic scalars describing the plasma, so that any transport coefficient is considered as a function of $(p,\bar\mu)$. Note that this is only a convenient choice of parameters.
From (\ref{diff1}), we then have
\be
\left(\partial D_{(s,t)}\over\partial p\right)_{\bar\mu} = {(s+1)\over\left(\epsilon+p\right)} D_{(s,t)}\quad,
\quad \left(\partial D_{(s,t)}\over\partial \bar\mu\right)_p = \xi_{(s,t)}\quad,\label{diff3}
\ee
and using the fact 
\be
\left(\partial T\over\partial p\right)_{\bar\mu} = {T\over \left(\epsilon+p\right)}\quad,
\ee
which can be derived by considering
\be
dp=s dT+n d\mu=s dT+ n d\left(T\bar\mu\right)= \left(s +n\bar\mu\right)dT+nTd\bar\mu ={\left(\epsilon+p\right)\over T} dT +n T d\bar\mu\quad,
\ee
the first equation gives us
\be
\left(\partial D_{(s,t)}\over\partial p\right)_{\bar\mu} =\left(s+1\right){1\over T}
\left(\partial T\over\partial p\right)_{\bar\mu}  D_{(s,t)}\quad,
\ee
which is solved by
\be
D_{(s,t)}=T^{s+1} \bar D_{(s,t)}(\bar\mu)\quad.\label{diff4}
\ee
Then, $\xi_{(s,t)}$ is given by the second equation of (\ref{diff3}) and
\be
\left(\partial T\over \partial\bar\mu\right)_p=-{n T^2\over\left(\epsilon+p\right)}\quad,
\ee
which can be obtained in a similar way.
The remaining step is to find $\bar D_{(s,t)}$. Inserting (\ref{diff4}) into (\ref{diff2}) and
using some of the above equations, one arrives at
\bear
&&\left(\partial\over\partial\bar\mu\right)_p\left(T^{s+1}\bar D_{(s,t)}\right)
-\left(s+1\right) T^s \left(\partial T\over\partial\bar\mu\right)_p \bar D_{(s,t)}
-2\left(N-s\right)T^{s+1} \bar D_{(s-1,t+1)} \nonumber\\
&&=T^{s+1} \left(\partial\bar D_{(s,t)}\over\partial\bar\mu\right) -2\left(N-s\right)T^{s+1} \bar D_{(s-1,t+1)}=0\quad,
\eear
which simply results in an iteration equation
\be
\left(\partial \bar D_{(s,t)}\over\partial\bar\mu\right)=2\left(N-s\right)\bar D_{(s-1,t+1)}\quad,
\ee
with an initial condition $\bar D_{(-1,N)}= {1\over 2N}\kappa\bar\mu$. This is easily solved as \footnote{We again neglect possible integration constants which are not fixed by the method.} 
\be
\bar D_{(s,t)}(\bar\mu) = {2^s (N-1)!\over (s+2)! (N-s-1) !} \kappa \bar\mu^{s+2}\quad.
\ee
From (\ref{diff4}) and the second equation of (\ref{diff3}), one finally has
\bear
D_{(s,t)}&=& {2^s (N-1)!\over (s+2)! (N-s-1) !} \kappa\,\,{\mu^{s+2}\over  T} \quad,\\
\xi_{(s,t)}&=& {2^s (N-1)!\over (s+1)! (N-s-1) !} \kappa \left(\mu^{s+1} - \left(s+1\over s+2\right) {n\over \left(\epsilon+p\right)} \mu^{s+2}\right)\quad. \label{f2}
\eear

\subsection{AdS/CFT correspondence}

We will close this section by confirming our results in the previous subsection in the AdS/CFT correspondence
via fluid/gravity computation \cite{Bhattacharyya:2008jc}. 
We start from the holographic bulk action in $(d+1)=(2N+1)$ dimensions,
\be
\left(16\pi G_{(d+1)}\right){\cal L} = R +d(d-1) -{1\over 4}F_{MN} F^{MN} -{C\over\sqrt{-g}}
\epsilon^{MP_1 Q_1\cdots P_N Q_N} A_M F_{P_1 Q_1}\cdots F_{P_N Q_N}\quad,
\ee
whose equations of motion are
\bear
R_{MN} +\left(d+{1\over 4(d-1)} F_{PQ}F^{PQ}\right) g_{MN} -{1\over 2} F_{MP} F_N^{\,\,\,\,P}&=&0\quad,
\nonumber\\
\nabla_P F^{MP} +{(d+2) C\over 2\sqrt{-g}} \epsilon^{M P_1 Q_1\cdots P_N Q_N} F_{P_1 Q_1}\cdots
F_{P_N Q_N}&=&0\quad.\label{eomlast}
\eear
The static charged black hole solution is
\be
ds^2=-r^2 V(r) dt^2 +2 dt dr +r^2 \sum_{i=1}^{d-1} \left(dx^i\right)^2\quad,\quad
A=-{q\over r^{d-2}} dt\quad,
\ee
with
\be
V(r)=1-{m\over r^d}+{(d-2) \over 2(d-1)}{q^2\over r^{2(d-1)}}\quad.
\ee
For fluid/gravity computation, the general boosted solution is
\be
ds^2_{(0)} = -r^2 V(r) u_\mu u_\nu dx^\mu  dx^\nu -2 u_\mu dx^\mu dr + r^2 \left(g_{\mu\nu}+u_\mu u_\nu\right)dx^\mu dx^\nu\quad,\quad A_{(0)}={q\over r^{d-2}} u_\mu dx^\mu\quad.
\ee
and one lets the parameters of the solution $(m(x),q(x),u^\mu(x))$ to vary in space-time, and 
systematically adds corrections to the above zero'th order metric in derivarive expansions of $(m(x),q(x),u^\mu(x))$
in order to satisfy the equations of motion (\ref{eomlast}).
For our case, we also need to introduce the external gauge potential in the derivative expansion, so we
in fact should extend the zero'th order bulk gauge field as
\be
A_{(0)}={q(x)\over r^{d-2} }u_\mu(x) dx^\mu + A_\mu(x)dx^\mu\quad,
\ee
where $A_\mu(x)$, by an abuse of notation, is understood as an external gauge potential.
The Ansatz is then
\be
ds^2= ds^2_{(0)}+\sum_k g_{MN}^{(k)}dx^M dx^N\quad,\quad A=A_{(0)}+A_M^{(k)}dx^M\quad,
\ee
where the corrections of $k$'th order $(g_{MN}^{(k)},A_M^{(k)})$
contain total $k$ number of derivatives of $(m(x),q(x),u^\mu(x),A_\mu(x))$.
It is convenient to use general coordinate/$U(1)$ gauge transformations to work in the gauge such that
\be
g_{rr}^{(k)}=0\quad,\quad g_{r\mu}^{(k)}\sim u_\mu\quad,\quad \sum_{i}g^{(k)}_{ii}=0\quad,\quad A^{(k)}_r=0 \quad.
\ee
At order $k$, the equations for $(g_{MN}^{(k)},A_M^{(k)})$ are simple second order linear ODE's along $r$
since $x^\mu$ derivatives of them are of higher order, and the sources for these linear ODE are given
in terms of solutions up to $(k-1)$'th order and their derivatives, and of order $k$ in total number of derivatives. It is an important fact that the linear ODE operators acting on $(g_{MN}^{(k)},A_M^{(k)})$
are universal, independent of $k$ and can thus be determined at the first order. See Ref.\cite{Bhattacharyya:2008jc} for details.

To find equations for $(g_{MN}^{(k)},A_M^{(k)})$ at the position, say $x^\mu=0$,
it is most convenient to work in the local rest frame where $u_\mu=(-1,\vec 0)$ at $x^\mu=0$,
and one can locally expand $(m(x),q(x),u^\mu(x),A_\mu(x))$ in powers of $x^\mu$'s up to the order of interest.
For example, at first order in derivatives one can use
\be
u_\mu(x)=\left(-1, x^\mu\partial_\mu u_i\right)\quad,\quad \left(m(x),q(x)\right)=(m,q)+x^\mu\left(\partial_\mu m,\partial_\mu q\right)\quad,\quad A_\mu(x)=-{1\over 2}F_{\mu\nu}x^\nu\quad.
\ee 
One can classify $(g_{MN}^{(k)},A_M^{(k)})$ according to their representations under the spatial symmetry group
 $SO(d-1)$ of our local rest frame, and since we are interested in the dissipative transverse $U(1)$ current, only
the vector modes $(g_{ti}^{(k)},A_i^{(k)})$ are relevant.

It is easy to check that the Chern-Simons term in the equations of motion (\ref{eomlast})
indeed starts appearing in the procedure at order $k=(N-1)$ in the vector mode equations due to
the total anti-symmetrization of $\epsilon$-tensor and the fact that the only non-zero zero'th order $F_{MN}^{(0)}$ is $F_{tr}^{(0)}$.
The linear differential operators along $r$ acting on $(g_{ti}^{(k)},A_i^{(k)})$ can be found
easily from Ref.\cite{Torabian:2009qk}. Since we are aiming to effects from anomaly or Chern-Simons term, we 
keep only source terms that come from the Chern-Simons term.
The relevant equations to solve are
\bear
\partial_r\left(r^{d+1}\partial_r\left(g_{ti}^{(N-1)}\over r^2\right)\right)+(d-2)q\left(\partial_r A_i^{(N-1)}\right)&=& 0\quad,\label{bulk1}\\
\partial_r\left(r^{d-1}V(r)\left(\partial_r A_i^{(N-1)}\right)\right) +(d-2)q \partial_r\left(g_{ti}^{(N-1)}\over r^2\right) &=&S_i(r)\quad,\label{bulk2}
\eear
where the source $S_i(r)$ coming from the Chern-Simons term is found to be
\be
S_i(r)=-{{1\over 2}d\left(d^2-4\right) C q}{1\over r^{d-1}}\epsilon^{i0j_1 k_1 \cdots j_{N-1}k_{N-1}} F^{(1)}_{j_1 k_1}(r)\cdots F^{(1)}_{j_{N-1} k_{N-1}}(r)\quad,
\ee
where
\be
F_{jk}^{(1)}(r)={q\over r^{d-2}}\left(\partial_j u_k - \partial_k u_j\right) +F_{ij}\quad.
\ee
Recalling $u_0=-1$ in our rest frame and expanding products of $F_{jk}^{(1)}$ in the source $S_i(r)$, one
easily recognizes $B_{(s,t)}^i$-structure appearing.
Explicitly, one has
\bear
S_i(r)={1\over 4}d^2\left(d^2-4\right)Cq \sum_{s} \,\left(2q\right)^s {(N-1)!\over s! (N-1-s)!} {1\over r^{s(d-2)+d-1}} B^i_{(s,t)}\,,\label{sif}
\eear

It is not difficult to solve (\ref{bulk1}) and (\ref{bulk2}) \cite{Torabian:2009qk};
first integrate (\ref{bulk2}) to get
\be
r^{d-1} V(r)\left(\partial_r A_i^{(N-1)}\right)+(d-2)q \left(g_{ti}^{(N-1)}\over r^2\right)
=\int_{r_H}^r dr'\,S_i(r') +{(d-2)q\over r_H^2} C_i\equiv -{r^{d-1}\over (d-2)q}I(r)\quad,\label{bulk3}
\ee
where considering the boundary condition at the horizon $r=r_H$, the integration constant $C_i$ is
equal to $g_{ti}^{(N-1)}(r_H)$ which should be determined later.
Inserting this into (\ref{bulk1}) removing $A_i^{(N-1)}$, one obtains a second order differential
equation for $g_{ti}^{(N-1)}$ which turns out to be after some manipulations
\be
\partial_r\left(r^{d+1}\left(V(r)\right)^2 \partial_r\left(g_{ti}^{(N-1)}\over r^2 V(r)\right)\right) =I(r)\quad,
\ee
which is readily integrated as
\be
g_{ti}^{(N-1)}(r)=r^2 V(r)\int^r_{\infty} dr'\,{1\over (r')^{d+1}\left(V(r')\right)^2}
\left(\int^{r'}_{r_H} dr''\,I(r'')-r_H^{d-1} V'(r_H) C_i\right)\quad,
\ee
where integration constants are fixed by considering regularity boundary conditions at the horizon.
Then, $A_i^{(N-1)}$ is given by integrating (\ref{bulk3}),
\be
A_i^{(N-1)}(r)= -\int^r_\infty dr'\,{1\over (r')^{d-1} V(r')}\left({(r')^{d-1}\over (d-2) q}I(r')+(d-2)q {g_{ti}^{(N-1)}(r')\over (r')^2}\right)\quad.
\ee
The above is the complete solution for $(g_{ti}^{(N-1)},A_i^{(N-1)})$ except $C_i$ still needs to be fixed.
The solution is regular for any $C_i$ and what fixes it is the Landau frame condition that $T_{ti}^{(N-1)}=0$
in our local rest frame, or equivalently $1\over r^{d-2}$-piece in the near boundary asymptotics of $g_{ti}^{(N-1)}$ should vanish.
This brings us the condition for $C_i$,
\be
\int^\infty_{r_H} dr\,I(r)= r_H^{d-1} V'(r_H) C_i\quad,
\ee
which finally determines $C_i$ as
\be
C_i=
-{(d-2)\over d}{q r_H^2\over m}
\int^\infty_{r_H}dr\,
{1\over r^{d-1}} \int^r_{r_H} dr'\, S_i(r')
\quad.\label{cif}
\ee
This completes the solution for $(g_{ti}^{(N-1)},A_i^{(N-1)})$.

Near the boundary $r\to\infty$, $A_i^{(N-1)}$ has the asymptotics,
\be
A_i^{(N-1)}(r)\to -{1\over (d-2)}\left(\int_{r_H}^\infty dr \,S_i(r) +{(d-2)q\over r_H^2} C_i\right) {1\over r^{d-2}}+{\cal O}\left({1\over r^d}\right)\quad,
\ee
and from this the $U(1)$ current from anomaly at $(N-1)$'th order is obtained as
\be
\nu_{i,{\rm anomaly}}^{(N-1)}= {(d-2)\over 16\pi G_{(d+1)}} \lim_{r\to\infty} r^{d-2}A_i^{(N-1)}(r)=
-{1\over 16\pi G_{(d+1)}}
\left(\int_{r_H}^\infty dr \,S_i(r) +{(d-2)q\over r_H^2} C_i\right)\quad.\label{u1}
\ee

There is one point we need to make clear;
in fact, the full boundary current gets an additional contribution from the variation of the bulk Chern-Simons term as was first stressed by Ref.\cite{Rebhan:2009vc},
\be
\nu^{(N-1)}_{i,{\rm full}}=\nu^{(N-1)}_{i,\rm{anomaly}}-{C d\over 16\pi G_{(d+1)}}\epsilon^{i\mu\nu_1\delta_1\cdots \nu_{N-1}\delta_{N-1}} A_\mu F_{\nu_1\delta_1}\cdots F_{\nu_{N-1}\delta_{N-1}}\quad.
\ee
However, rigorous holographic renormalization of the theory in Ref.\cite{Sahoo:2010sp} shows that what enters in the energy-momentum Ward identity
\be
\partial_\mu T^{\mu\nu} = F^{\mu\nu}j_\nu\quad,\label{ward}
\ee
is the current $j_\nu$ obtained only by the near boundary asymptotics like $\nu^{(N-1)}_{i,\rm{anomaly}}$ in the above, not the full current $j_{\nu,{\rm full}}$ including additional contribution from the Chern-Simons term \footnote{We thank A.Yarom for discussions on this.}. The basic reason for this is that the Chern-Simons term is topological and does not couple to the metric at all, so that it gives no contribution to the energy-momentum tensor. The choice of a  current one is working with does not matter as long as one is clear about its definition and the correct Ward identities. Since our discussions are based on the energy-momentum Ward identity of the form (\ref{ward}), the correct current we have to use here is $\nu^{(N-1)}_{i,\rm{anomaly}}$. 

Finally, we need to fix the coefficient $C$ in front of the Chern-Simons term, and the point in the above paragraph is also relevant here. Recall that our basic hydrodynamic equations (\ref{basic2})
assume the energy-momentum Ward identity of the form (\ref{ward}), so that we need to match $C$ such that
the divergence of the current obtained only from the near boundary asymptotics, not of the full current, agrees with the second equation of (\ref{basic2}). 
The bulk equation of motion of the $U(1)$ gauge field (\ref{eomlast}) near the boundary $r\to\infty$ dictates that 
\be
\partial_\mu j^\mu = {(d+2) C\over 32\pi G_{(d+1)}}
\epsilon^{\mu_1\nu_1\cdots\mu_{N}\nu_{N}}F_{\mu_1\nu_1}\cdots F_{\mu_{N}\nu_{N}}\quad,
\ee
where $j^\mu$ in the above is the current obtained only from the near boundary asymptotics as in (\ref{u1}).
Comparing with (\ref{basic2}), we can fix $C$ as
\be
C={64\pi G_{(d+1)} \over d^2 (d+2)} \kappa\quad.\label{fixc}
\ee

With (\ref{sif}), (\ref{cif}), (\ref{u1}), and (\ref{fixc}), one finally obtains after some algebra 
\be
\nu^{(N-1)}_{i,\rm{anomaly}}=\sum_s\, \xi_{(s,t)} B^i_{(s,t)}\quad,
\ee
where
the expressions for $\xi_{(s,t)}$ is
\bear
\xi_{(s,t)}={ 2^s (N-1)!\over (s+1)!(N-s-1)!}{\kappa q^{s+1}\over r_H^{(s+1)(d-2)} }\left(1-\left(s+1\over s+2\right)\left(d-2\over d\right) {q^2\over m r_H^{d-2}}\right)\quad.
\eear
Using the relations
\be
\epsilon+p= d p = {d m\over 16\pi G_{(d+1)}}\quad,\quad n={(d-2) q\over 16\pi G_{(d+1)}}\quad,\quad
\mu={q\over r_H^{d-2}}\quad,
\ee
it is easily checked that this agrees precisely with our result (\ref{f2}) in the previous subsection.

\vskip 1cm \centerline{\large \bf Acknowledgement} \vskip 0.5cm
We thank Y. Burnier, J. Ellis, J. Liao, Y. Oz, E. Shuryak, D. Son, D. Teaney, A.~Yarom and I.~Zahed  for useful discussions. D.K. is grateful to CERN Theory Division for hospitality during the completion of this work. This research was supported by the U.S. Department of Energy under Contracts No.
DE-FG-88ER40388, DE-AC02-98CH10886, and DE-FG-88ER41723.

 \vfil

\end{document}